\documentclass{aastex62}

\usepackage{graphicx}
\usepackage{latexsym}
\usepackage{amsfonts,amsmath,amssymb}
\usepackage{epstopdf}
\usepackage{natbib}
\usepackage{mathrsfs}
\usepackage{algorithm}
\usepackage[utf8]{inputenc}
\usepackage{hyperref}
\usepackage[noend]{algpseudocode}
\usepackage{xcolor}

\usepackage{csquotes}

\def\apjref#1;#2;#3;#4 {\par\pni\ #1,  #2, {\bf #3}, #4. \par}

\shortauthors{Gaches, Offner \& Bisbas}

\begin{document}
\title{The Astrochemical Impact of Cosmic Rays in Protoclusters I: Molecular Cloud Chemistry}

\author[0000-0003-4224-6829]{Brandt A.L. Gaches}
\affiliation{Department of Astronomy, University of Massachusetts - Amherst, Amherst, MA 01003, USA}
\email{bgaches@astro.umass.edu}

\author[0000-0003-1252-9916]{Stella S.R. Offner}
\affiliation{Department of Astronomy, The University of Texas at Austin, Austin, TX 78712, USA}
\email{soffner@astro.as.texas.edu}

\author[0000-0003-2733-4580]{Thomas G. Bisbas}
\affiliation{I.~Physikalisches Institut, Universit\"at zu K\"oln, Z\"ulpicher Stra{\ss}e 77, Germany}
\affiliation{Department of Physics, Aristotle University of Thessaloniki, GR-54124 Thessaloniki, Greece}
\email{bisbas@ph1.uni-koeln.de; tbisbas@auth.gr}

\correspondingauthor{Brandt Gaches}

\begin{abstract}
We present astrochemical photo-dissociation region models in which cosmic ray attenuation has been fully coupled to the chemical evolution of the gas. We model the astrochemical impact of cosmic rays, including those accelerated by protostellar accretion shocks, on molecular clouds hosting protoclusters. Our models with embedded protostars reproduce observed ionization rates. We study the imprint of cosmic ray attenuation on ions for models with different surface cosmic ray spectra and different star formation efficiencies. We find that abundances, particularly ions, are sensitive to the treatment of cosmic rays. We show the column densities of ions are under predicted by the ``classic'' treatment of cosmic rays by an order of magnitude. We also test two common chemistry approximations used to infer ionization rates. We conclude that the approximation based on the H$_3^+$ abundance under predicts the ionization rate except in regions where the cosmic rays dominate the chemistry. Our models suggest the chemistry in dense gas will be significantly impacted by the increased ionization rates, leading to a reduction in molecules such as NH$_3$ and causing H$_2$-rich gas to become [C II] bright.
\end{abstract}


\section{Introduction}\label{sec:intro}
Molecular cloud dynamics and chemistry are sensitive to the ionization fraction. The chemistry of molecular clouds is dominated by ion-neutral reactions \citep{watson1976} and thus controlled by the ionization fraction. The gas (kinetic) temperature of a typical molecular cloud with an average H-nucleus number density of $n\approx10^3\,{\rm cm}^{-3}$ is approximately 10~K for cosmic-ray ionization rates $\zeta\lesssim10^{-16}\,{\rm s}^{-1}$ \citep{bisbas2015,bisbas2017}, thus rendering neutral-neutral reactions inefficient. Ionization in molecular clouds is produced in three difference ways: UV radiation, cosmic rays (CRs), and X-Ray radiation. Ultra-violet radiation, from external O- and B-type stars and internal protostars, does not penetrate very far into the cloud due to absorption by dust. However, cosmic rays, which are relativistic charged particles, travel much further into molecular clouds and dominate the ionization fraction when $A_V \geq 5\,{\rm mag}$ \citep{mckee1989,strong2007,grenier2015}. CR-driven chemistry is initiated by ionized molecular hydrogen, ${\rm H}_2^+$ \citep{dalgarno2006}. The ion-neutral chemistry rapidly follows:
\begin{equation*}
{\rm CR + H_2 \rightarrow H_2^+ + e^- + CR'}
\end{equation*}
\begin{equation*}
{\rm H_2^+ + H_2 \rightarrow H_3^+ + H,}
\end{equation*}
where CR$'$ is the same particle as CR but with a lower energy. The ejected electron from the first reaction can have an energy greater than the ionization potential of H$_2$ and thus cause further ionization. Once H$_3^+$ forms, more complex chemistry follows, thereby 
creating a large array of hydrogenated ions:
\begin{equation*}
{\rm X + H_3^+ \rightarrow [XH]^+ + H_2}.
\end{equation*}

Both HCO$^+$ and N$_2$H$^+$, important molecules used to map the dense gas in molecular clouds, form this way with X being CO and N$_2$, respectively. These species are also used to constrain the cosmic ray ionization rate (CRIR) \citep[i.e.,][]{caselli1998, ceccarelli2014}. OH$^+$ and H$_n$O$^+$ are also formed this way through H$_3^+$ and H$^+$ \citep{hollenbach2012}. In addition, at low column densities ($A_V<1\,{\rm mag}$), which is typical of the boundaries of molecular clouds), the non-thermal motions between ions and neutrals may overcome the energy barrier of the reaction 
\begin{equation*}
{\rm C^+ + H_2 \rightarrow CH^+ + H,}
\end{equation*}
leading to an enhancement of the CO column density \citep{federman1996, visser2009} and a shift of the H{\sc i}-to-H$_2$ transition to higher $A_V$ \citep{bisbas2019}.

The ionization fraction controls the coupling of the magnetic fields to the gas, influencing non-ideal magnetohydrodynamic (MHD) effects such as ambipolar diffusion \citep{mckee2007}. These non-ideal effects can play a signficant role in the evolution in the cores and disks of protostars. On galactic scales, numerical simulations have shown that CRs can help drive large outflows and winds out of the galaxy \citep[e.g.,][]{girichidis2016}. Our study focuses on the impact of CRs on Giant Molecular Cloud scales which is typically not resolved fully in such simulations.

There have been a plethora of studies modeling the impact of CRs on chemistry and thermal balance \citep[i.e.,][]{caselli1998, bell2006, meijerink2011, bayet2011, clark2013, bisbas2015}. However, in these studies, and the vast majority of astrochemical models, the CRIR is held constant throughout the cloud, despite the recognition that CRs are attenuated and modulated while traveling through molecular gas \citep{schlickeiser2002, padovani2009, schlickeiser2016, padovani2018}. Galactic-CRs, thought to be accelerated in supernova remnants or active galactic nuclei, are affected by hadronic and Coulombic energy losses and screening mechanisms that reduce the flux with increasing column density \citep{strong2001, moskalenko2005, evoli2017}. The modulation of CRs has not previously been included within astrochemical models of molecular clouds due to the difficulty in calculating the attenuation and subsequent decrease in the CRIR \citep{wakelam2013, cleeves2014}.

Given that CRs are thought to be attenuated, it is expected that the ionization rate should decline within molecular clouds. However, recent observations do not universally show a lower ionization rate. \cite{favre2017} inferred the CRIR towards 9 protostars and found a CRIR consistent with the rate inferred for galactic CRs. The OMC-2 FIR 4 protocluster, hosting a bright protostar, is observed to have a CRIR 1000 higher than the expected rate from galactic CRs \citep{ceccarelli2014, fontani2017, favre2018}. \cite{gaches2018b} show that this system can be modelled assuming the central source is accelerating protons and electrons within the accretion shocks on the protostar's surface. In general, accreting, embedded protostars may accelerate enough CRs to cancel the effect of the attenuation of external CRs at high column densities, producing a nearly constant ionization rate throughout the cloud \citep{padovani2016, gaches2018b}.
 
Typical accretion shocks and shocks generated by protostellar jets satisfy the physical conditions necessary to accelerate protons and electrons \citep{padovani2016, gaches2018b}. Accretion shocks in particular are a promising source of CRs since they are strong, with velocities exceeding 100 km/s and temperatures of millions of degrees Kelvin \citep{hartmann2016}. \cite{gaches2018b} calculated the spectrum of accelerated protons in protostellar accretion shocks and the attenuation through the natal core assuming that the CRs free-stream outwards. These models suggest that clusters of a few hundred protostars accelerate enough CRs into the surrounding cloud to exceed the ionization rate from Galactic CRs. 

In this study, we explore the effects of protostellar CRs on molecular cloud chemistry by employing the model of \cite{gaches2018b}. We implement an approximation for CR attenuation into the astrochemistry code {\sc 3d-pdr} \citep{bisbas2012} to account for CR ionization rate gradients. We investigate the signatures of a spatially varying ionization rate. We further explore the impact of protostellar CR sources and their observable signatures.

The layout of the paper is as follows. In \S\ref{sec:methods} we present the CR and protostellar models and describe the implementation of CR attenuation into {\sc 3d-pdr}. We discuss our results in \S\ref{sec:res}. Finally, in \S\ref{sec:disc} we create observational predictions and compare them to observations.

\section{Methods}\label{sec:methods}

\subsection{Protocluster Model}\label{sec:proto_model}

We generate protoclusters following the method of \cite{gaches2018a}, where the model cluster is parameterized by the number of stars and gas surface density, N$_*$ and $\Sigma_{\rm cl}$, respectively. These parameters are connected to the star formation efficiency $\varepsilon_g = M_*/M_{\rm gas}$, where M$_{\rm gas}$ is related to $\Sigma_{\rm cl}$ following \cite{mckee2003} $\Sigma_{\rm cl} = \frac{M_{\rm gas}}{\pi R^2}$, where the cloud radius, $R$, is determined from the density distribution (See \S\ref{sec:dens}). We model protoclusters with surface densities in the range $0.1 \leq \Sigma_{\rm cl} \leq 10$ g cm$^{-2}$ and with a number of stars in the range $10^2 \leq N_* \leq 10^4$.  In this parameter space, we always consider $\varepsilon_g \leq 25\%$.

We generate $N_{\rm cl} = 20$ protoclusters for each point in the parameter space and adopt the average CR spectrum for the chemistry modelling. We use the Tapered Turbulent Core (TTC) accretion history model \citep{mckee2003, offner2011}, where the protostellar core is supported by turbulent pressure and accretion is tapered to produce smaller accretion rates as the protostellar mass, $m$, approaches the final mass, $m_f$. \cite{gaches2018a} showed the TTC model is able to reproduce the bolometric luminosities of observed local protoclusters.
  
\subsection{Cosmic Ray Model}\label{sec:cr}

We brielfy summarize the CR acceleration and propagation model in \cite{gaches2018b} and refer the reader to that paper for more details. We assume CRs are accelerated in the accretion shock near the protostellar surface. Accreting gas is thought to flow along magnetic field lines in collimated flows with the shock at the termination of the flow. We assume the shock velocities are comparable to the free-fall velocity at the stellar surface. Following \cite{hartmann2016}, we assume fully ionized strong shocks with the shock front perpendicular to the magnetic field lines. We adopt a mean molecular weight $\mu_I = 0.6$ and a filling fraction of accretion columns on the surface, $f = 0.1$.

We calculate the CR spectrum under the Diffusive Shock Acceleration (DSA) limit, also known as first-order Fermi acceleration \cite[e.g., reviewed in][]{drury1983, kirk1994, melrose2009}. Under DSA, the CR momentum distribution is a power-law, $f(p) \propto p^{-q}$, where $q$ is related to the shock properties. We attenuate the CR spectrum through the protostellar core following \cite{padovani2009}. \cite{padovani2018} presented updated attenuation models for surface densities up to 3000 g cm$^{-2}$, but the results remain unchanged for the surface of concern in this work. The core surface density and radius for a turbulence-supported core are \citep{mckee2003}:
\begin{subequations}
\begin{align}
\Sigma_{\rm core} &= 1.22\Sigma_{\rm cl} = 0.122 {\rm \, g \, cm^{-2}} \left (\frac{\Sigma_{\rm cl}}{{\rm 0.1 \, g \, cm^{-2}}} \right )\\
N({\rm H_2})_{\rm core} &= \frac{\Sigma_{\rm core}}{\mu_M m_H} \\
 &= 3\times 10^{22} {\rm \, cm^{-2}\, } \left ( \frac{\Sigma_{\rm core}}{0.122 {\rm \, g \, cm^{-2}}} \right ) \left ( \frac{2.4}{\mu} \right ) \nonumber\\
R_{\rm core} &= 0.057 \Sigma_{\rm cl}^{-\frac{1}{2}} \left ( \frac{m_f}{30 ~M_{\odot}} \right )^{\frac{1}{2}} {~~~\rm pc} \\
&= 0.104 {\rm \, pc} \left ( \frac{\Sigma_{\rm cl}}{0.1 {\rm \, g \, cm^{-2}}} \right )^{-\frac{1}{2}} \left ( \frac{m_f}{10 {\rm \, M_{\odot}}}\right )^{\frac{1}{2}} \nonumber ,
\end{align}
\end{subequations}
where $\mu_M = 2.4$ is the mean molecular weight for a molecular gas. We calculate the total protocluster CRIR by summing over the $N_*$ attenuated 
CR spectra.

\subsection{Density Structure}\label{sec:dens}
We assume the molecular cloud density is described by an inverse power law
\begin{equation}
n(r) = n_0 \left ( \frac{R}{r} \right )^2,
\end{equation}
where $R$ is the cloud radius and $n_0$ is the number density with an inner radius of 0.1 pc. The $r^{-2}$ dependence matches the solution for isothermal collapse \citep{shu1977}. We take $n_0 = 100$ cm$^{-3}$, corresponding to a gas regime in which the cloud is expected to be mostly molecular under typical conditions. The radius is set by constraining the total surface density by $\Sigma_{\rm cl}$ as defined:
\begin{equation}
\int_{R_c}^R n(r) dr = \frac{\Sigma_{\rm cl}}{\mu_M m_H},
\end{equation}
where $R_c$ is the inner radius delineating the transition between the molecular cloud and protostellar core. The turbulent-linewidth, $\sigma$, of a cloud with density profile  $n(r) \propto r^{-2}$ and a virial $\alpha$ parameter is \citep{bertoldi1992}:
\begin{equation}
\sigma = \left ( \frac{G M^2 \alpha}{3 \mu_M m_H \bar{n} V R} \right )^{\frac{1}{2}} ,
\end{equation}
where $\bar{n}$ is the volume-averaged density from $n(r)$, $G$ is the gravitational constant, and $V = \frac{4}{3} \pi R^3$ is the volume of the molecular cloud. We take $\alpha = 2$ for our fiducial model \citep{heyer2015}.

\subsection{Chemistry with Cosmic Ray Attenuation}
We use a modified version of the {\sc 3d-pdr} astrochemistry code\footnote{The code can be downloaded from \url{https://uclchem.github.io}, including the new modifications presented in this paper.} introduced in \cite{bisbas2012}. {\sc 3d-pdr} solves the chemical abundance and thermal balance in one-, two- and three-dimensions for arbitrary gas distributions. The code can be applied to arbitrary three dimensional gas distributions, such as post-processing simulations \citep{offner2013, offner2014, bisbas2018}. Here, we use the code in one dimension to model a large parameter space. We adopt the \cite{mcelroy2013} {\sc umist12} chemical network containing 215 species and approximately 3,000 reactions. We assume the gas is initially composed of molecular H$_2$ with the rest being atomic with abundances from \cite{sembach2000} shown in Table \ref{tab:abund}. Cooling is included from line emission, which is mainly due to carbon monoxide at low temperatures and forbidden lines of [OI],[CI] and [CII] at higher temperatures. Heating is due to dissipation of turbulence,  photoelectric heating of dust from far-ultraviolet emission,  H$_2$ fluorescence and CR heating of gas. We use a temperature floor of 10 K. Previously, {\sc 3d-pdr} included CRs via a single global CRIR parameter. See \cite{bisbas2012} for more technical details.

\begin{deluxetable}{cc}
\tablecolumns{2}
\tablecaption{Atomic Abundances \label{tab:abund}}
\tablehead{\colhead{Species} & \colhead{Abundance Relative to H}}
\startdata
H & 1.0\\
He & 0.1\\
C & 1.41$\times10^{-4}$\\
N & 7.59$\times10^{-5}$\\
O & 3.16$\times10^{-4}$\\
S & 1.17$\times10^{-5}$\\
Si & 1.51$\times10^{-5}$\\
Mg & 1.45$\times10^{-5}$\\
Fe & 1.62$\times10^{-5}$
\enddata
\tablecomments{Atomic abundances adopted from \cite{sembach2000}.}
\end{deluxetable}

We modify {\sc 3d-pdr} to account for CR attenuation through the cloud. Currently, our implementation is restricted to one-dimensional models where we assume spherical symmetry. {\sc 3d-pdr} calculates the CRIR from $N_{\rm SRC}$ CR sources. The user provides a CR spectrum for any number of sources and whether it is external (incident at the surface) or internal (originating at the cloud center). In 1D, the fluxes are defined on either surface of the domain. The flux due to external sources is attenuated while the point sources are assumed to radiate isotropically; both are attenuated and spatially diluted. The spectra are attenuated after every update of the molecular abundances to keep the amount of H$_2$ for interaction losses self-consistent. {\sc 3d-pdr} stores the initial CR flux in $N_{\rm ene}$ bins and self-consistently calculates $\zeta$ from all sources across the domain. Point sources require a user-set radial scaling factor, $r_s$, and a transport parameter, $a$. For our model results, we set $r_s = R_C$ to represent the core radius. Point source CR spectra, $j(E, r)$, are attenuated by the H$_2$ column density \citep{padovani2009} and diluted by the radial distance following
\begin{equation}
j(r) \propto \left ( \frac{r_s}{r + r_s}\right )^a,
\end{equation}
to approximate the effects of transport. Solving the transport equations for Galactic transport problems has been done with specialized codes, such as {\sc Galprop} \citep{moskalenko1998} and {\sc Dragon2} \citep{evoli2017}. Fully solving the steady-state transport equations are beyond the scope of this work but will be investigated in the future.

In our study, we include two CR flux sources. First, we include the internal protostellar clusters discussed above. We set the radial scaling $r_s = R_C = 0.1$ pc, which is approximately the size of a protostellar core. Second, we include an external isotropic CR flux to model interstellar CRs. We follow \cite{ivlev2015} and parameterize the external flux as
\begin{equation}
j_{\rm ext} = C \frac{E^{\alpha}}{\left ( E + E_0 \right )^{\beta}} {~~~\rm (particles ~eV^{-1} ~cm^{-2} ~s^{-1} ~sr^{-1})}.
\end{equation}
We use their ``low'' model ($\mathcal{L}$), with $C = 2.4\times10^{15}$, $E_0 = 650$ MeV, $\alpha = 0.1$ and $\beta = 2.8$ and their ``high'' model ($\mathcal{H}$), with $C = 2.4\times10^{15}$, $E_0 = 650$ MeV, $\alpha = -0.8$ and $\beta = 1.9$. The $\mathcal{L}$ model is a direct extrapolation of the Voyager-1 data \citep{stone2013}, while the $\mathcal{H}$ is a maximal model to correct for any possible effects of the solar magnetic field. The CRIR is calculated by integrating the spectrum multiplied by the H$_2$ cross section:
\begin{equation}
    \zeta_p = 2 \pi \int j(E) \sigma_{i}(E) dE ,
\end{equation}
where the factor of $2\pi$ accounts for irradiating the 1-D surface on one side and $\sigma_i(E)$ is the H$_2$ ionization cross section with relativistic corrections \citep{krause2015}. The code allows for an arbitrary number of energy bins, $N_{\rm bins}$, for input CR spectrum. We compared the CRIR for bin sizes ranging from N$_{\rm bin} = 4$ to $1000$ and found that N$_{\rm bins} > 40$ only produces changes in the CRIR at the 1\% level. We do not fully solve for primary or secondary electrons. Therefore, we multiply the proton CRIR by $\frac{5}{3}$ to account for the electron population \citep{dalgarno1958, takayanagi1973}.

Our fiducial parameters for the study are shown in Table \ref{tab:params}. Table \ref{tab:models} shows the full suite of models we adopt. The model names describe the included physics: L/H denotes using the $\mathcal{L}$/$\mathcal{H}$ (Low/High) external spectrum, NI denotes no internal sources, DI denotes internal sources with $a = 1$ (diffusive transport), RI denotes internal sources with $a = 2$ (rectilinear transport) and NA denotes no internal sources or CR attenuation. We study the impact of these parameters in Section \ref{sec:res}. 

\begin{deluxetable}{ccc}
\tablecolumns{3}
\tablecaption{Model Parameters \label{tab:params}}
\tablehead{\colhead{Parameter} & Definition & \colhead{Fiducial Value}}
\startdata
$\mu_{\rm I}$ & Reduced gas mass & 0.6 (ionized) \\
$\mu_{\rm M}$ & Reduced gas mass & 2.4 (neutral molecular) \\
$n_0$ & Density at edge of cloud & $10^2$  cm$^{-3}$ \\ 
$N_{\rm src}$ & Number of CR Sources & 2 \\
$N_{\rm bins}$ & Number of CR spectrum bins & 40 \\
a & CR transport parameter & 1 \\
$r_s$ & Scaling radius for CR flux & 0.1 pc \\
$\alpha$ & Cloud virial parameter & 2 \\
\enddata
\end{deluxetable}

\begin{deluxetable}{c|cccc}
\tablecolumns{5}
\tablecaption{Models examined. L/H denotes Low/High external spectrum, NI denotes no internal sources, DI denotes internal sources with $a = 1$ (diffusive transport), RI denotes internal sources with $a = 2$ (rectilinear transport) and NA denotes no internal sources or CR attenuation. \label{tab:models}}
\tablehead{\colhead{Name} & \colhead{Source Transport} & \colhead{Internal} & \colhead{External Field} & \colhead{Attenuation} }
\startdata
LDI\label{model:fid} & $r^{-1}$ & $\checkmark$ & $\mathcal{L}$ & $\checkmark$ \\
LRI\label{model:rec} & $r^{-2}$ & $\checkmark$ & $\mathcal{L}$ & $\checkmark$ \\
LNI\label{model:ni} & ... & ...  & $\mathcal{L}$ & $\checkmark$ \\
LNA\label{model:na} & ... & ...  & $\mathcal{L}$ & ...  \\
HNI\label{model:hni} & ... & ... & $\mathcal{H}$ & $\checkmark$ \\
HDI\label{model:hdi} & $r^{-1}$ & $\checkmark$ & $\mathcal{H}$ & $\checkmark$ \\
\enddata
\end{deluxetable}

\section{Results}\label{sec:res}

\subsection{Cosmic Ray Spectrum}\label{sec:crsp}
\begin{figure}
	\centering
    \includegraphics[width=0.5\textwidth]{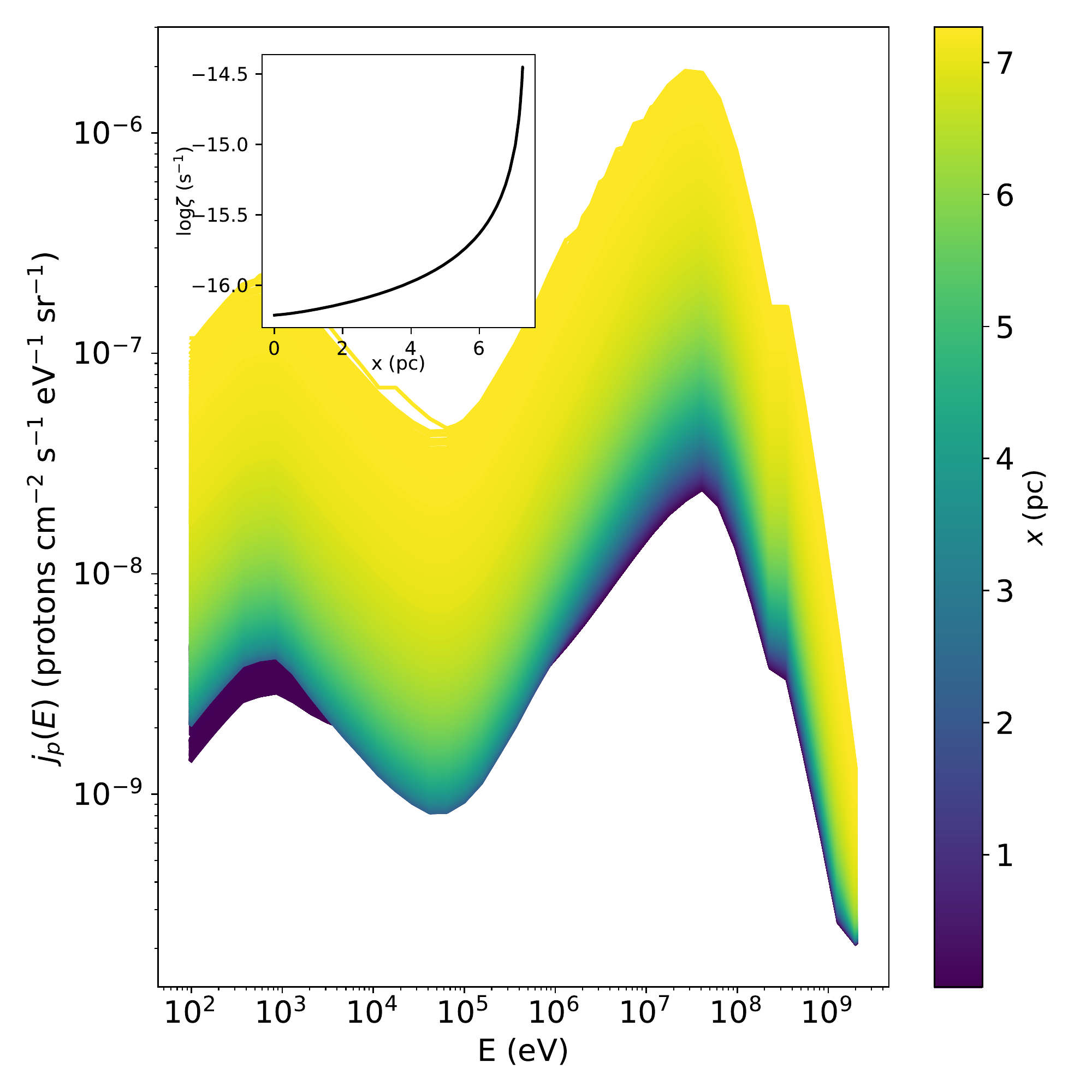}
	\caption{\label{fig:crspecs} Proton cosmic ray spectrum with line color indicating position within the cloud for $\Sigma_{\rm cl} = 0.75$ g cm$^{-2}$ and N$_*$ = 750 using the LDI CR model. Inset: Cosmic ray ionization rate versus position, $x$, into the cloud, where $x = 0$ is the cloud surface.}
\end{figure}

Our modified {\sc 3d-pdr} code requires as an input the flux of CR protons for any number of sources. As a result, the CR proton flux and CRIR throughout the spatial domain become outputs rather than inputs. Figure \ref{fig:crspecs} shows an example CR spectrum for a molecular cloud with $\Sigma_{\rm cl} = 0.75$ g cm$^{-2}$ and $N_* = $ 750 using the LDI model. The CR proton flux increases inside the cloud because of the embedded sources. The double peaked shape of the spectrum is due to peaks in the loss function from ionization and Coulomb losses. The inset shows the CRIR as a function of the position within the cloud. In this model, the ionization rate climbs nearly two orders of magnitude throughout the cloud with increasing proximity to the protostellar cluster.

\subsection{Cosmic Ray Ionization Rate Models}\label{sec:crir}
A number of prescriptions have been used to calculate the CRIR from observed column densities of various tracer species \citep{caselli1998, indriolo2012}. The inclusion of CR attenuation allows us to directly test the accuracy of these approximations.  Our astrochemical models provide the abundances throughout the cloud and the local CRIR in-situ. We test two different prescriptions that are typically used for diffuse and dense gas, respectively, from \cite{indriolo2012}. The first, and simplest, denoted as ``Simple Electron Balance'' (SEB), estimates the CRIR using only the abundance of H$_3^+$ and $e^-$:
\begin{equation}\label{eq:ebal}
    \zeta = k_e n({\rm e^-}) \frac{n({\rm H_2})}{n({\rm H_3^+})},
\end{equation}
where $k_e$ is the H$_3^+$ electron-recombination rate and $n({\rm e^-})$, $n({\rm H_2})$ and $n({\rm H_3^+})$ are the densities of electrons, molecular hydrogen and H$_3^+$, respectively. The second approximation includes the destruction of H$_3^+$ with CO and O, which we denote the ``Reduced Analytic'' (RA) model:
\begin{equation}\label{eq:simpanaly}
    \zeta = \frac{x({\rm H_3^+})}{x({\rm H_2})} n_H \left [ k_e x({\rm e^-}) + k_{\rm CO} x({\rm CO}) + k_O x({\rm O}) \right ]  \times \left [ 1 + \frac{2 k_3 x({\rm e^-})}{k_2 f({\rm H_2})} + \frac{2 k_4}{k_2} \left ( \frac{1}{f({\rm H_2})} - 1 \right ) \right ],
\end{equation}
where $k_i$ are the reaction rate coefficients for the reactions in Table from \cite{indriolo2012} (repeated in Table \ref{tab:react} below), $x_i$ is the abundance of species i with respect to total Hydrogen nuclei and $f({\rm H_2}) = 2n({\rm H_2})/n_H$ is the molecular hydrogen fraction. 

\begin{figure*}
	\centering
    \begin{tabular}{c}
    \includegraphics[width=0.85\textwidth]{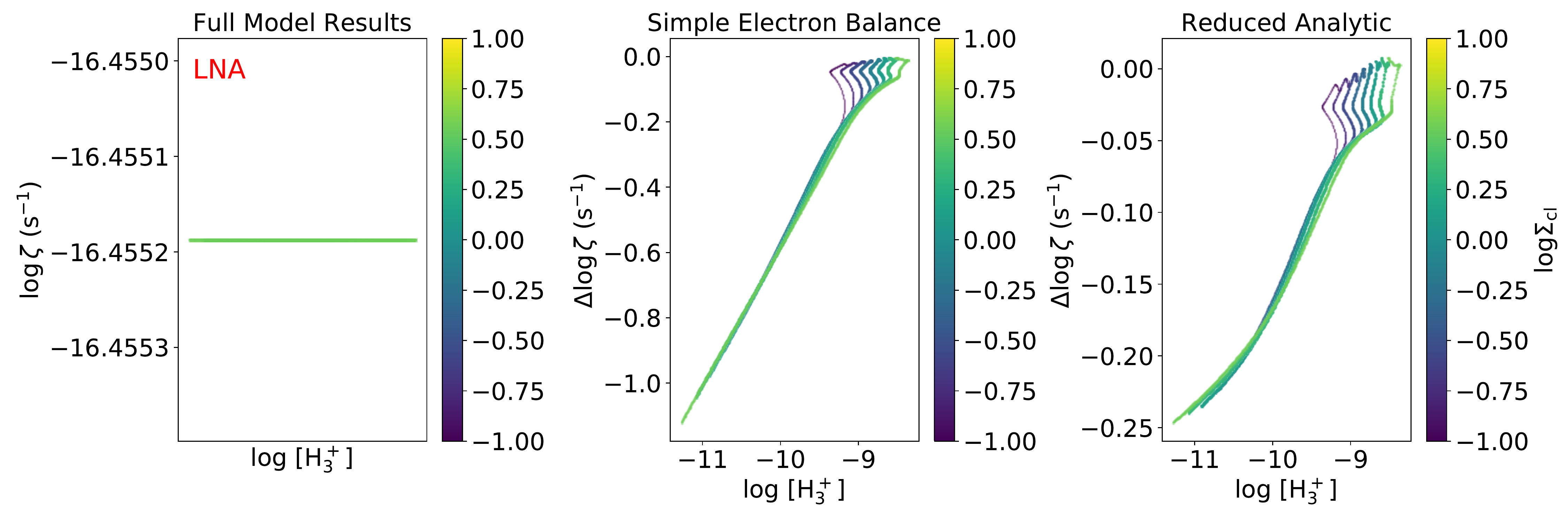} \\ \hline
    \includegraphics[width=0.85\textwidth]{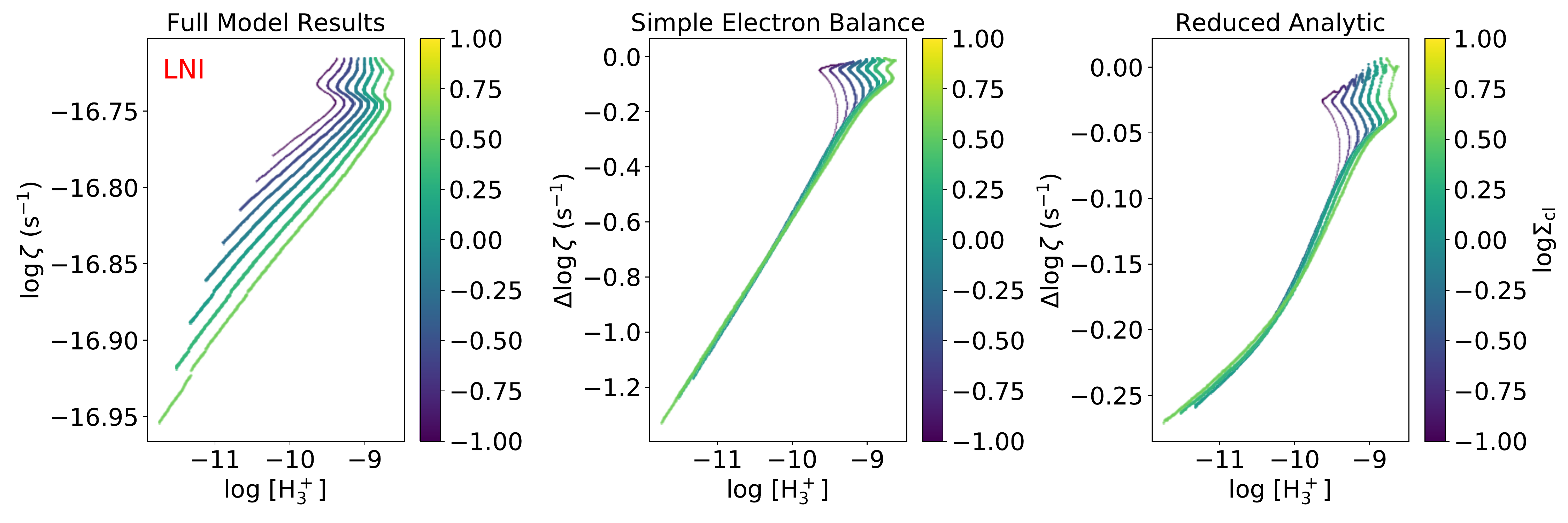} \\ \hline
    \includegraphics[width=0.85\textwidth]{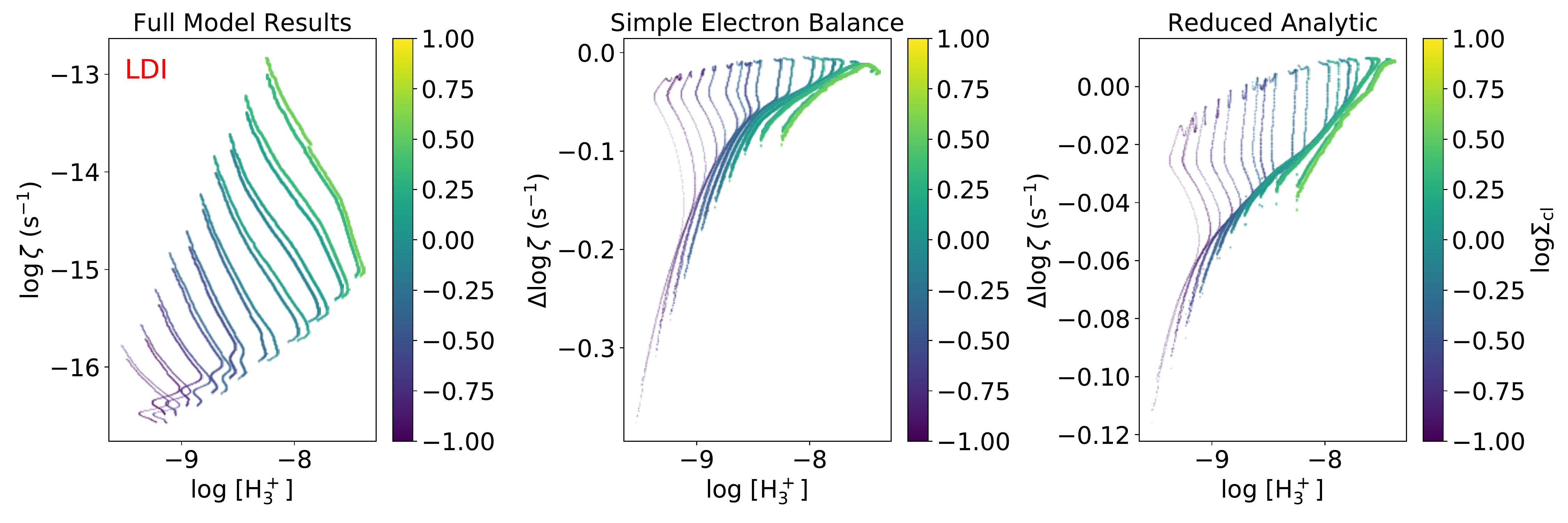} \\ \hline
    \includegraphics[width=0.85\textwidth]{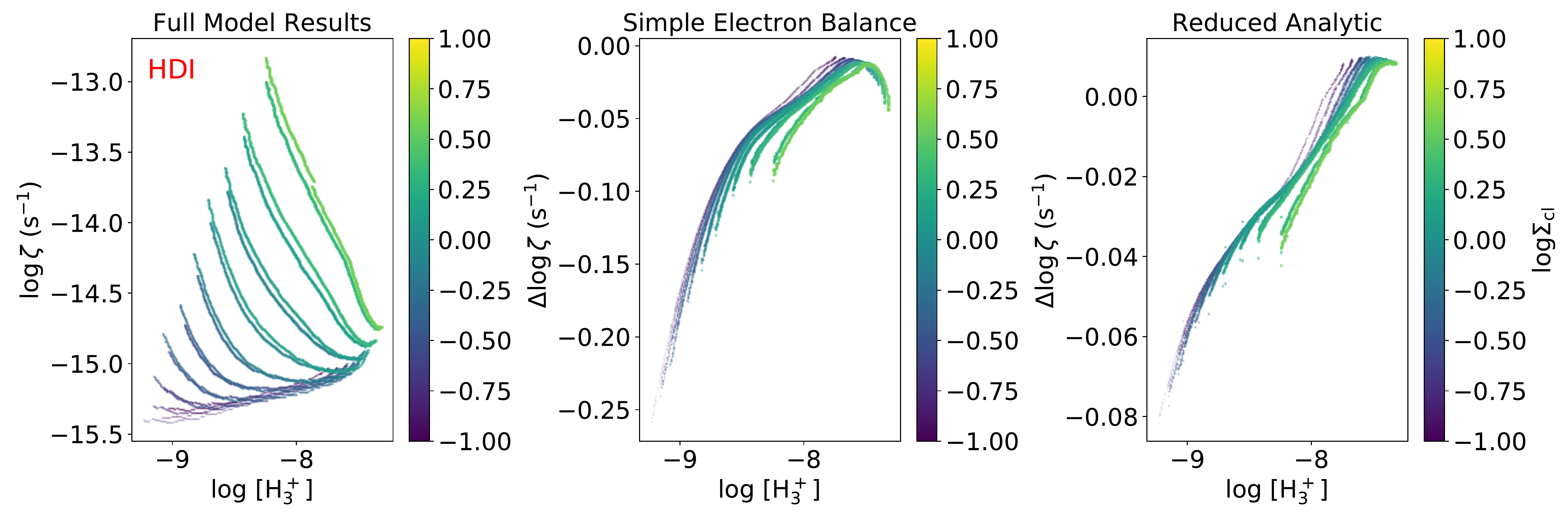} \\
    \end{tabular}
	\caption{\label{fig:zeta} Left: Cosmic ray ionization rate as a function of H$_3^+$ abundance using the full astrochemical model results. Middle: Relative logarithmic error in $\zeta$, $\Delta \log \zeta$, calculated using  electron-balance approximation (Equation \ref{eq:ebal}) and full astrochemical model. Right: Same as middle row but using the reduced analytic approximation (Equation \ref{eq:simpanaly}). Color: Gas surface density, $\Sigma_{\rm cl}$ in g cm$^{-2}$. Models in descending order from the top are: LNA, LNI, LDI, HDI.}
\end{figure*}

Figure \ref{fig:zeta} shows the calculated CRIR using the full model and the approximations in Equations \ref{eq:ebal} and \ref{eq:simpanaly} as a function of the H$_3^+$ abundance. We show the cases of four different CR models: LNA, LNI, LDI and HDI. The first model, LNA, is of particular importance since it represents the simplest one-dimensional PDR model. Observations typically assume 0D distribution, such that the ratio of column densities is equal to the ratio of the abundances. This makes a tacit assumption that the ionization rate is the same throughout the domain. We find that both approximations produce a large range of CRIRs -- even when the input CRIR is fixed due to other effects impacting the chemistry, notably the influence of the external FUV radiation. {\it The SEB and RA approximation models systematically underestimate the CRIR and produce an artificial spread in the inferred CRIR.} When internal sources are included, we find that both approximations infer the CRIR reasonably well. When there are no internal sources, both approximations underestimate the CRIR by up to an order of magnitude and in general do not represent any real spread in the CRIR accurately.

\begin{deluxetable*}{ccc}
\tablecolumns{3}
\tablecaption{Reactions for Reduced Analytic H$_3^+$ Chemistry}\label{tab:react}
\tablehead{\colhead{Reaction} & \colhead{Rate Coefficient (cm$^3$ s$^{-1}$)} & \colhead{Reference}}
\startdata
H$_2^+$ + H$_2$ $\rightarrow$ H$_3^+$ + H & $k_2 = 2.09 \times 10^{-9}$ & \cite{theard1974} \\
H$_2^+$ + e$^-$ $\rightarrow$ H + H & $k_3 = 1.6\times10^{-8} (T/300)^{-0.43}$ & \cite{mitchell1990} \\
H$_2^+$ + H $\rightarrow$ H$_2$ + H$^+$ & $k_4 = 6.4\times10^{-10}$ & \cite{karpas1979} \\
H$_3^+$ + e$^-$ $\rightarrow$ products & $k_5 = k_e = -1.3\times10^{-8} + 1.27 \times 10^{-6} T_e^{-0.48}$ & \cite{mccall2004} \\
H$_3^+$ + CO $\rightarrow$ H$_2$ + HCO$^+$ & $k_6 = 1.36\times10^{-9}(T/300)^{-0.142} \exp{3.41/T}$ & \cite{klippenstein2010} \\
H$_3^+$ + CO $\rightarrow$ H$_2$ + HCO$^+$ & $k_7 = 8.49\times10^{-10}(T/300)^{0.0661}\exp{-5.21/T}$ & \cite{klippenstein2010} \\
H$_3^+$ + O $\rightarrow$ H$_2$ + OH$^+$ & $k_8 = k_O = 1.14\time10^{-9}(T/300)^{-0.156}\exp{-1.41/T}$ & \cite{klippenstein2010} \\
\enddata
\tablecomments{Equations \ref{eq:ebal} and \ref{eq:simpanaly} reactino rates where $k_{\rm CO} = k_6 + k_7$ in Equation \ref{eq:simpanaly}. We omit the Nitrogen reaction from \cite{indriolo2012} Table 3 since it is not in use by either approximation.}
\end{deluxetable*}

\subsection{Impact of Cosmic Ray Sources on Cloud Chemistry}\label{sec:chem}

\begin{figure*}
	\plotone{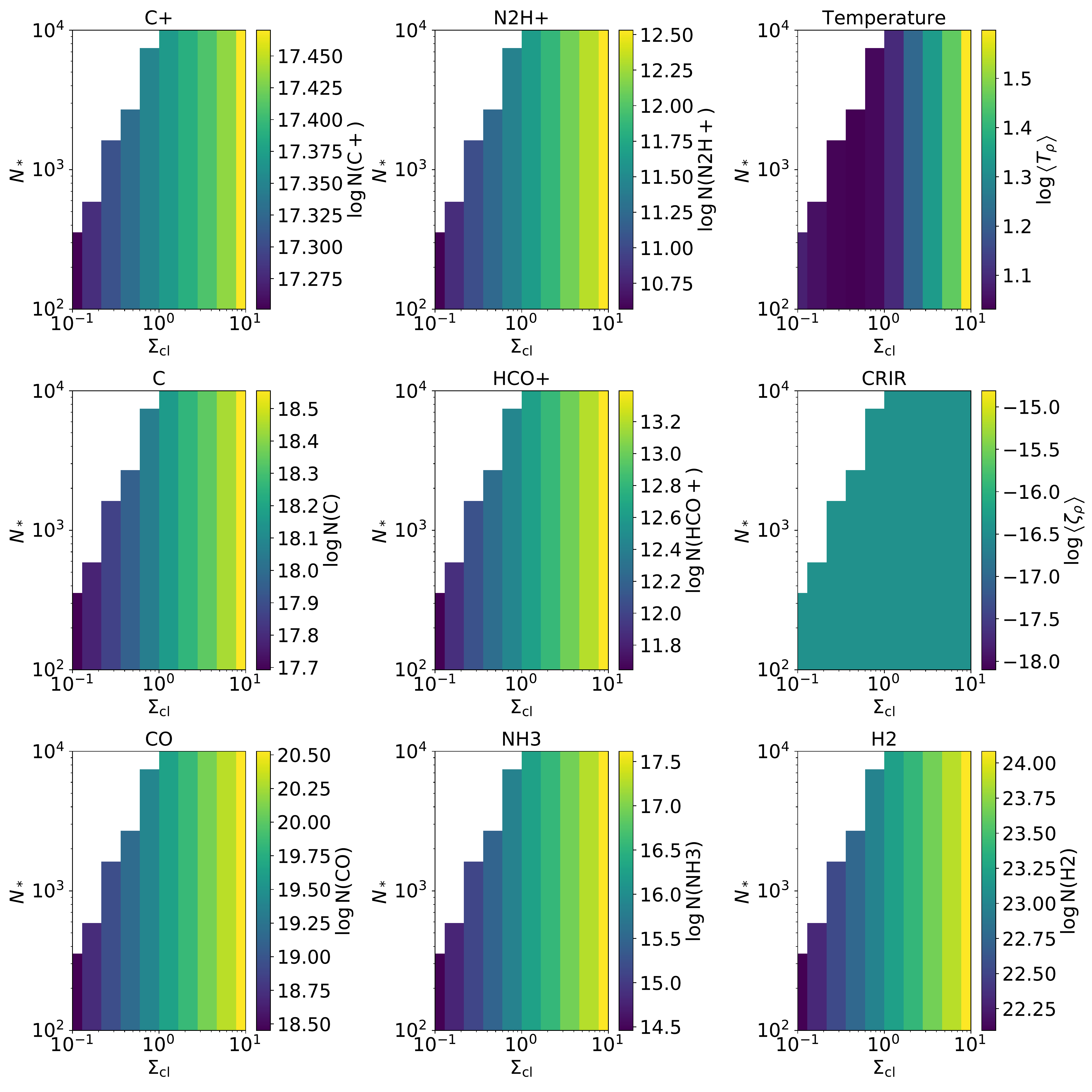}
    \caption{\label{fig:colVirNA}Column density of different molecular species as a function of the number of stars in the protocluster, $N_*$ and the mass surface density, $\Sigma_{\rm cl}$ for the LNA model. The last 3 panels on the far right show the density averaged temperature, CRIR and the total gas column density.}
\end{figure*}

\begin{figure*}
	\plotone{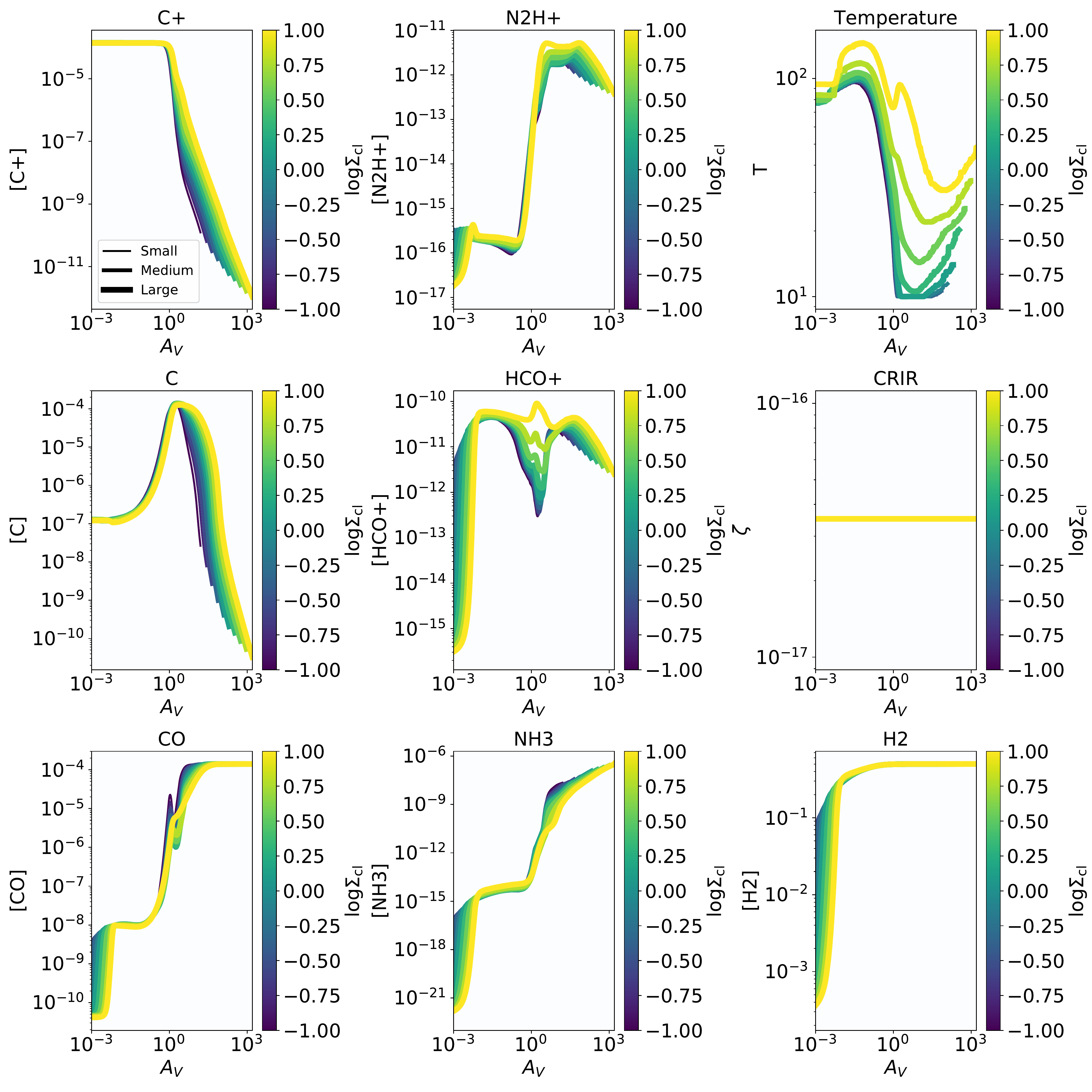}
    \caption{\label{fig:profVirNA} The abundances, $[ {\rm X} ]$, for various molecular species, as a function of the visual extinction into the cloud, $A_V$, for the LNA model. The top two panels on the right column show the temperature and CRIR as a function of the $A_V$. The colorbar indicates the gas surface density, $\Sigma_{\rm cl}$. The line width indicates the number of protostars in the cluster with ``Small'' = $10^2$, ``Medium'' = $10^3$ and ``Large'' = $10^4$ protostars.}
\end{figure*}

We examine in detail two different CR models: the traditional LNA and the LDI model. Figure \ref{fig:colVirNA} shows the column densities of different species and density averaged temperature and CRIR for the LNA model as a function of $\Sigma_{\rm cl}$ and $N_*$. The total column density of each species increases with the gas surface density, $\Sigma_{\rm cl}$, and hence the total gas mass. Furthermore, we find across the whole parameter space that N(CO) $>$ N(C) $>$ N(C$^+$). This qualitative behavior is to be expected with no internal sources. Figure \ref{fig:profVirNA} shows the abundance profiles for the LNA model as a function of $A_V$ into the cloud, $\Sigma_{\rm cl}$ and $N_*$. Since there are no embedded sources in this model, there is no difference between models of different $N_*$. The abundance profiles for C$^+$, C and CO exhibit the expected ``layered'' behavior \citep{draine2011}: C$^+$ is confined to the surface, C exists in a thin, warm layer and CO asymptotically approaches an abundance of [CO/H$_2$] $\approx 10^{-4}$. Similarly the abundance of NH$_3$ steadily increases into the cloud. 

The abundance ratio ${\rm [HCO^+/N_2H^+]}$ is sometimes used to infer the CRIR under the assumption the two molecules are co-spatial \citep[i.e.,][]{ceccarelli2014}. We find that, while they share some local maxima,  they are not completely co-spatial {in agreement with} the turbulent cloud study of \cite{gaches2015}.Moreover, observations show that while they are not entirely co-spatial, there is overlap in the emission regions \citep[i.e.,][]{jorgensen2004,ceccarelli2014, storm2016, favre2017, pety2017, pound2018}. In particular, we show HCO$^+$ can exist at much lower $A_V$ than N$_2$H$^+$. Due to similar critical densities however, the two molecules thermalize at nearly the same densities. 

\begin{figure*}
	\plotone{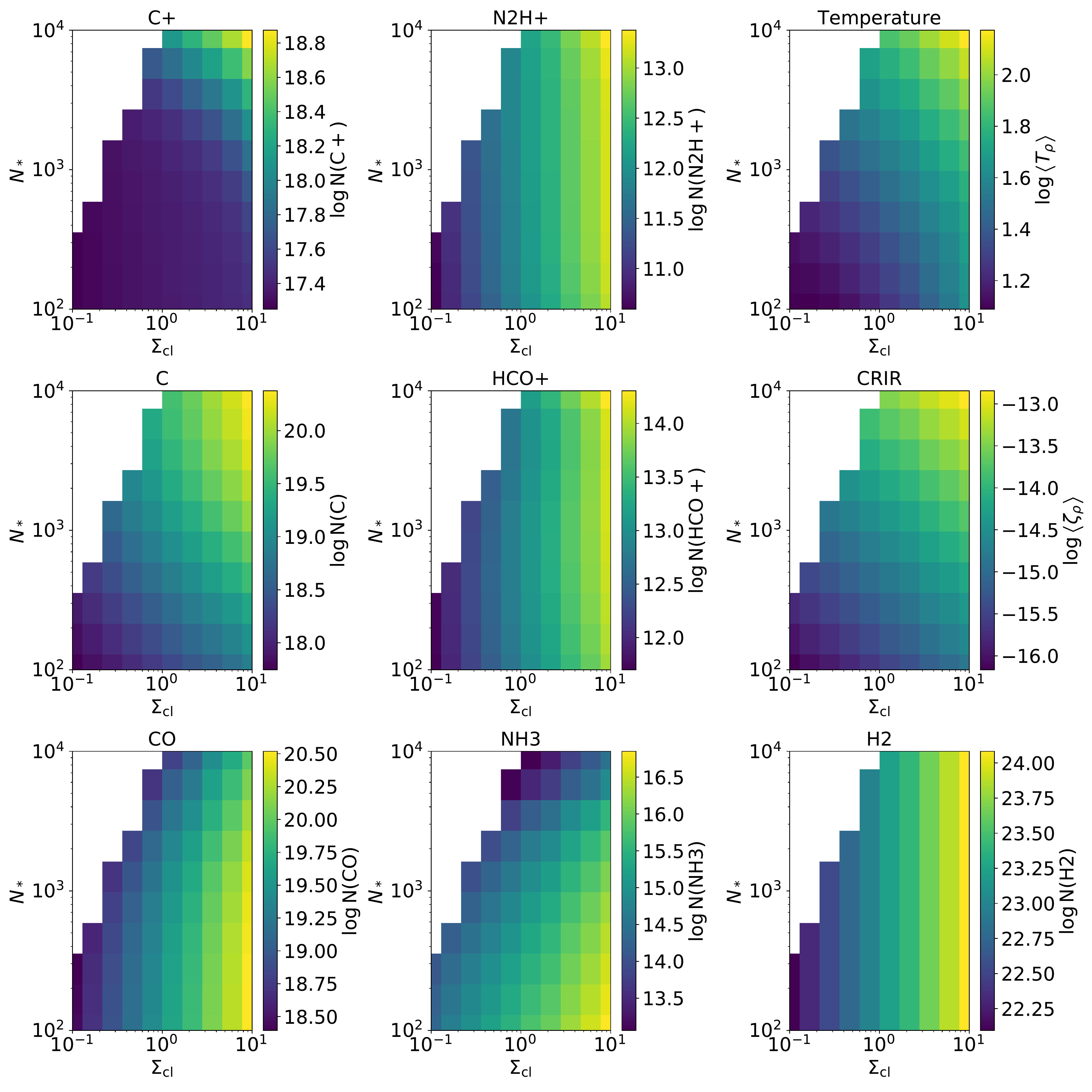}
    \caption{\label{fig:colVir}Same as Figure \ref{fig:colVirNA} but for model LDI.}
\end{figure*}

Figure \ref{fig:colVir} shows the column densities across the parameter space for the LDI model. Here we find a very different behavior compared to the LNA model shown in Figure \ref{fig:colVirNA}, where the differences are especially pronounced for the more diffuse gas tracers. The column densities are no longer strictly functions of $\Sigma_{\rm cl}$ but depend on $N_*$. For large, massive star-forming regions (upper right corner in each panel), the gas becomes CO deficient and C rich while the bulk of gas remains molecular. Similarly, there is a slight increase in the column density of HCO$^+$ and N$_2$H$^+$ due to the increase in ionization. The qualitative trends exhibited by C$^+$, C, and marginally by HCO$^+$ and N$_2$H$^+$, follow that of the density-averaged CRIR, $\langle \zeta_\rho \rangle$. 

\begin{figure*}
	\plotone{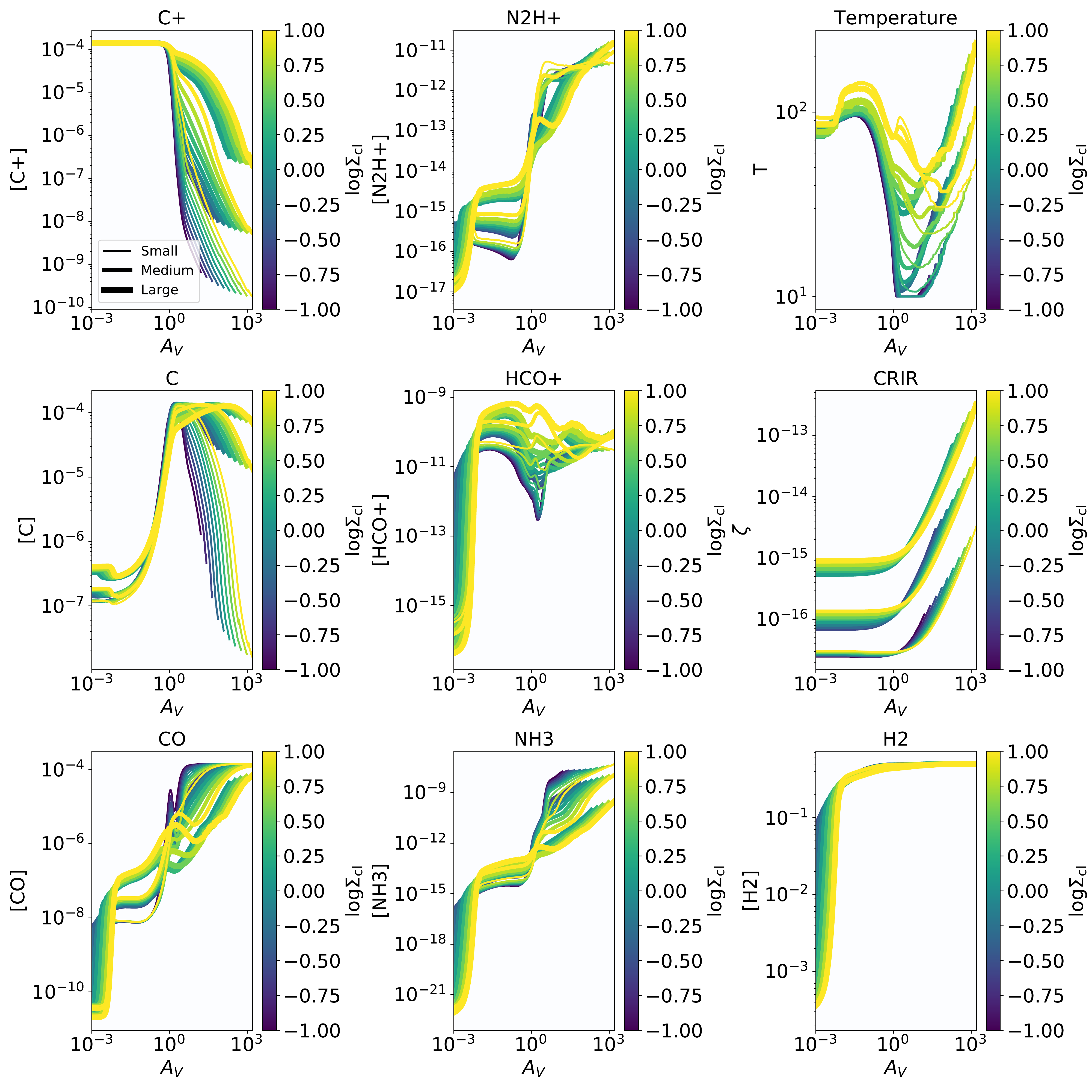}
    \caption{\label{fig:profVir} Same as Figure \ref{fig:profVirNA} but for model LDI.}
\end{figure*}

The effect of an embedded protocluster is also visible in the abundance profiles. Figure \ref{fig:profVir} shows the abundance profiles for the LDI model. We find that CO only approaches abundances of $10^{-4}$ for clusters with little embedded star formation.For smaller mass clouds (i.e., smaller $\Sigma_{\rm cl}$), the C$^+$, C and CO abundance remains fairly unchanged compared to the LNA model. In the most massive clouds, the amount of CO at $A_V \leq 1$ is enhanced by an order of magnitude and reduced by an order of magnitude at $A_V \geq 5$. We further see a reduction in N$_2$H$^+$ at mid-$A_V$ with an enhancement of HCO$^+$. Likewise the gas temperature exceeds $T > 30$ K for most of the clouds with $\Sigma_{\rm cl} > 0.25$ g cm$^{-2}$. As $A_V \rightarrow 10^3$, the differences between the molecular ion abundances is much less due to the greatly increased density compared to the surface of the cloud. The abundance of H$_2$ in the dense gas is unaffected by the increased CRIR.

\begin{figure*}
    \centering
    \includegraphics[width=\textwidth]{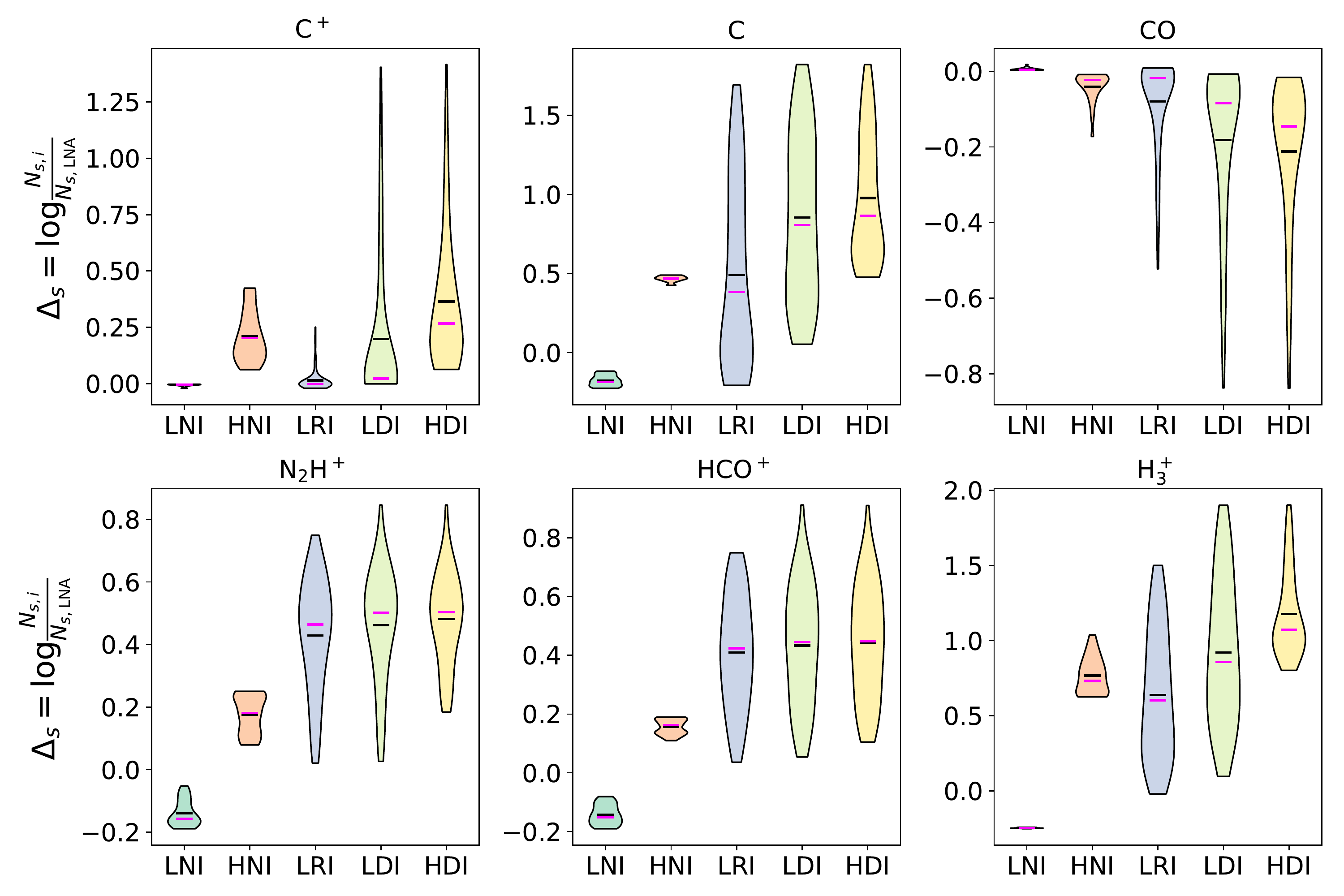}
    \caption{\label{fig:violins} Logarithmic difference distribution, $\Delta_s = \log \frac{N_{\rm s,i}}{N_{\rm s, LNA}}$ comparing the difference in column density for molecular species, $s$, for five different cosmic ray models, $i \subset $ (LNI, HNI, LRI, LDI, HDI), compared to the LNA model. Distributions are generated using the model clouds across the ($\Sigma_{\rm cl}$ - $N_*$) parameter space. The black and magenta line indicate the mean and median, respectively.}
\end{figure*}

We now statistically quantify the impact of different CR models on the six different molecules: C$^+$, C, CO, N$_2$H$^+$, HCO$^+$ and H$_3^+$. We investigate the H$_3^+$ column density because it is the simplest molecule that can be used to constrain the CRIR \citep{dalgarno2006}. We calculate the column density logarithmic difference:
\begin{equation}
\Delta_s = \log \frac{N_{s,i}}{N_{s, {\rm LNA}}},
\end{equation}
with $s$ representing the different species, and $i$ the different CR models exluding the LNA model. Figure \ref{fig:violins} shows violin plots representing the probability distribution of $\Delta_s$ using all clouds in the $(\Sigma_{\rm cl}-N_*)$ space. In all cases, CO is never enhanced but rather depleted. This is because the maximum abundance [CO] = 10$^{-4}$ is set by the C/O ratio. Our models generally increase the local CRIR, thereby dissociating the CO and reducing its abundance. We find very little difference between the LNI and LNA for all molecules except for N$_2$H$^+$ and HCO$^+$, which exhibit a 25\% linear dispersion. This is caused by the impact of higher ionization rates towards the surface of the clouds. The HNI model,which has the highest overall CRIR at the surface, shows a clear offset for the atomic and ionic species and a slight deficit for CO. In models LRI, LDI and HDI there is a significant dispersion in the column density difference, $\Delta_s$, in all species. Figure \ref{fig:violins} demonstrates that considerable care must be taken when modeling observed column densities of atomic or ionic species: the possible error, $\Delta_s$, in the modeled column densities may be off by an order of magnitude depending on the transport of the cosmic rays and the amount of ongoing embedded star formation. The CRIR is not the only factor that leads to the creation of molecular ions, as typically assumed in observational studies. The abundances are influenced by the FUV flux, which is also enhanced by a central protocluster \citep{gaches2018a}.

\subsection{Abundance Ratio Diagnostics for the Cosmic Ray Ionization Rate}

Line and abundance ratios of various tracers are often used to constrain the CRIR in dense gas. The species are typically assumed to be co-spatial (although as demonstrated in Section \ref{sec:chem} that is not typically the case). We examine two different ratio diagnostics: global diagnostics using column densities and local diagnostics using the local abundance ratios and CRIRs.

\begin{figure*}
	\centering
    \begin{tabular}{c}
    \includegraphics[width=\textwidth]{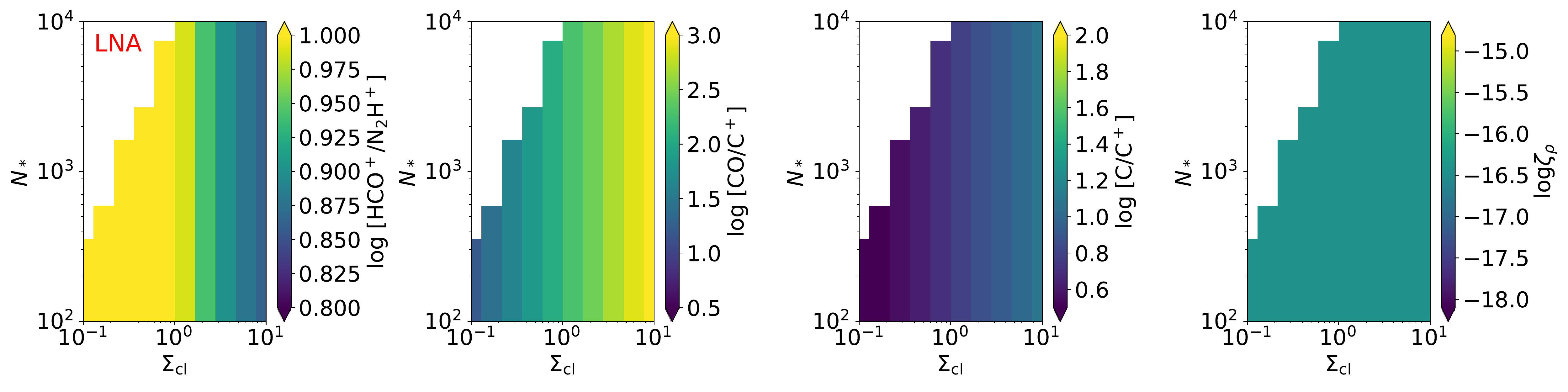} \\ \hline
    \includegraphics[width=\textwidth]{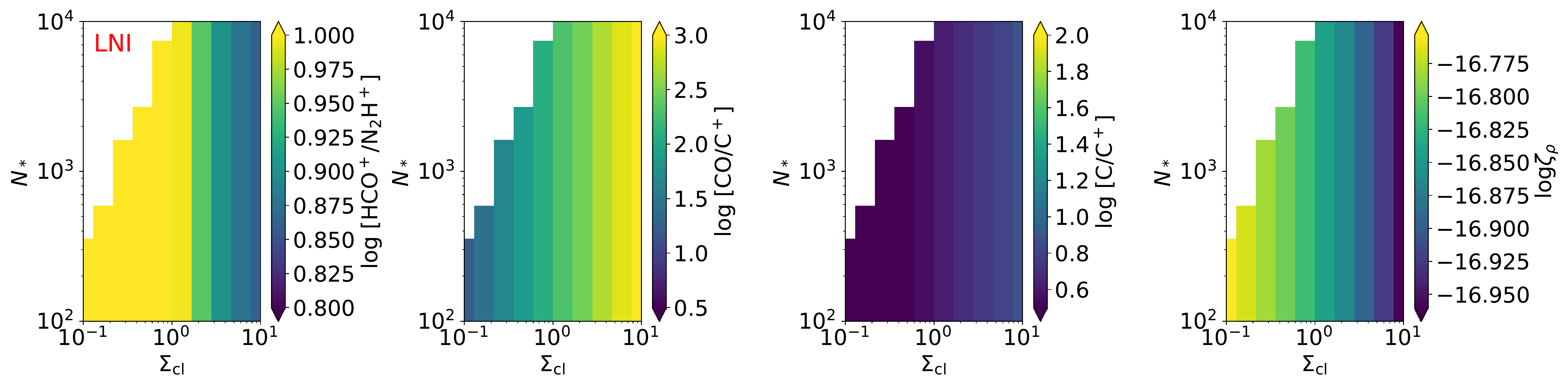} \\ \hline
    \includegraphics[width=\textwidth]{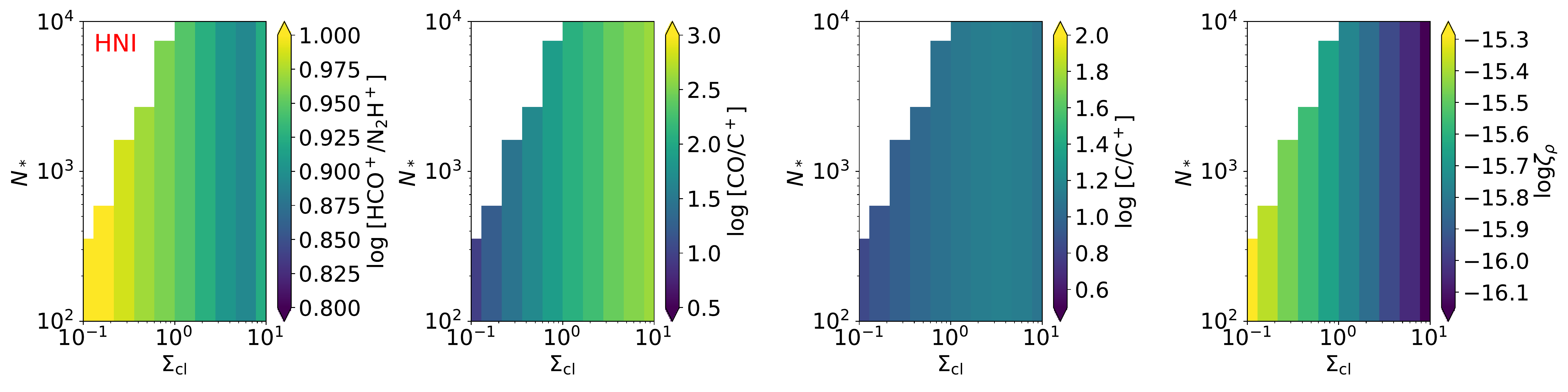} \\ 
    \end{tabular}
	\caption{\label{fig:ratsI}Abundance ratios versus  $\Sigma_{\rm cl}$ and $N_*$. White contours and labels show the star formation efficiency $\varepsilon = M_*/M_{\rm gas}$. Far left: [HCO$^+$]/[N$_2$H$^+$]. Middle left: [CO/C$^+$]. Middle right: [C/C$^+$]. Far right: Density-weighted average cosmic ray ionization rate, $<\zeta_\rho>$. The models are in descending order from top: LNA, LNI, HNI. We use the same scales for the individual ratios in Figures \ref{fig:ratsI} and \ref{fig:ratsS} to facilitate comparison between models. 
	}
\end{figure*}

Figure \ref{fig:ratsI} shows three different column density ratios for the LNA, LNI and HNI models: ${\rm [HCO^+/N_2H^+]}$, ${\rm [CO/C^+]}$ and ${\rm [C/C^+]}$. We find that the column density ratios in these cases increase monotonically with $\Sigma_{\rm cl}$. The ratio of ${\rm [HCO^+/N_2H^+]}$ is nearly constant, changing by less than a factor of two across two dex of $\Sigma_{\rm cl}$. The HNI case shows a slightly different behavior with a slight local minimum in ${\rm [HCO^+/N_2H^+]}$ at $\Sigma_{\rm cl} = 6$ g cm$^{-2}$. The trends in these models are not due to changes in the CRIR but rather in the total amount of gas column. The [CO/C$^+$] ratio shows a buildup of CO compared to C$^+$. This is to be expected in an externally irradiated model: C$^+$ remains consistently on the surface, while the amount of CO continues to build with $\Sigma_{\rm cl}$ with a proportional increase in the amount of dense gas. Similarly, the [C/C$^+$] remains fairly constant since these species exist only in limited areas of $A_V$.

\begin{figure*}
	\centering
    \begin{tabular}{c}
    \includegraphics[width=\textwidth]{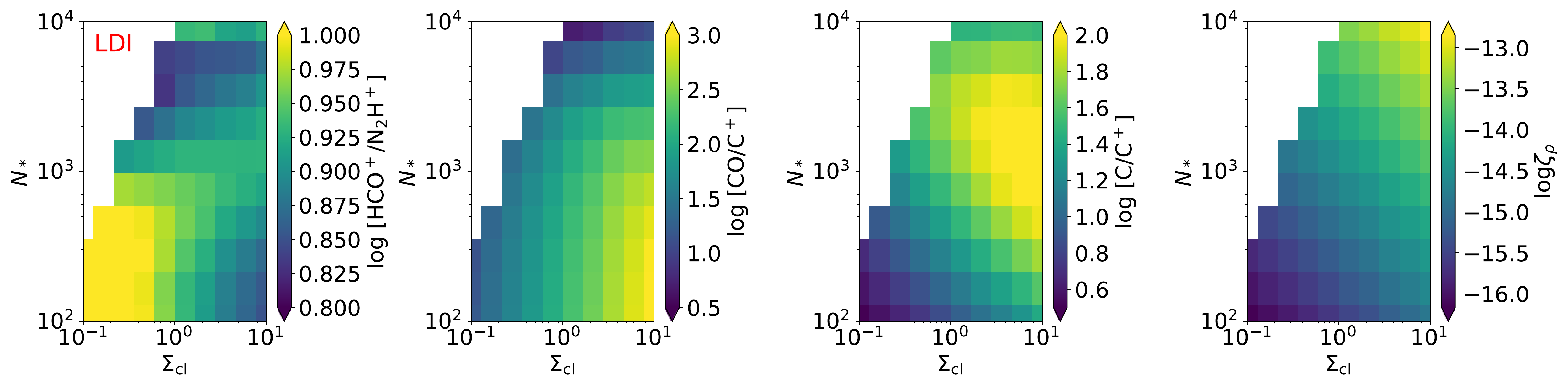} \\ \hline
    \includegraphics[width=\textwidth]{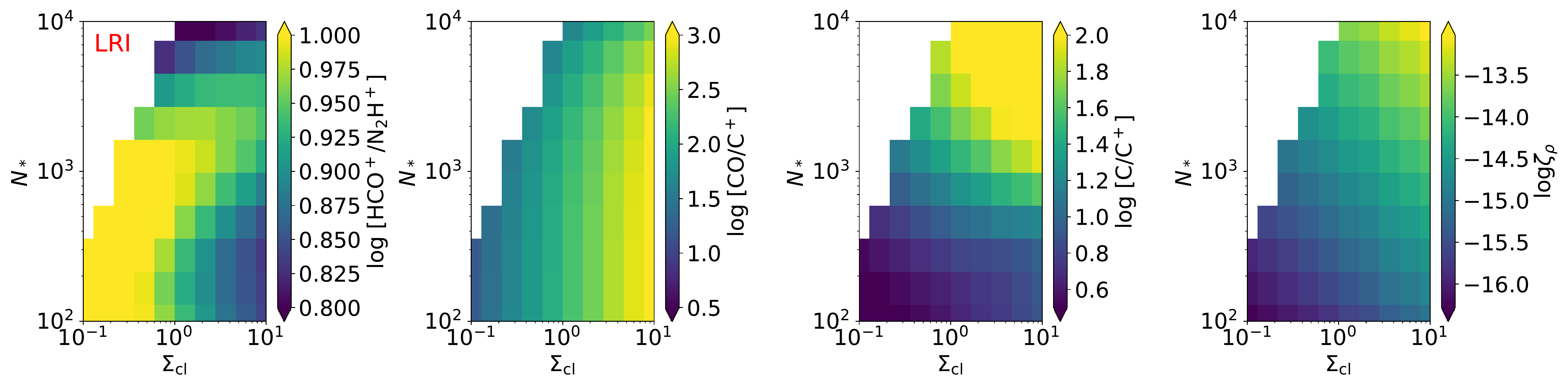} \\ \hline
    \includegraphics[width=\textwidth]{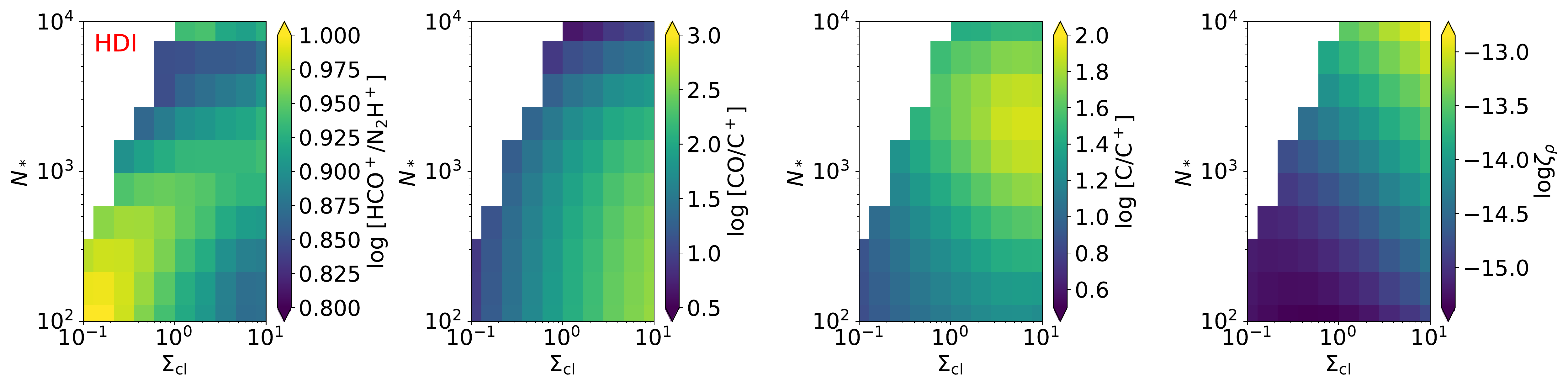} \\
    \end{tabular}
	\caption{\label{fig:ratsS}Abundance ratios versus  $\Sigma_{\rm cl}$ and $N_*$. White contours and labels show the star formation efficiency $\varepsilon = M_*/M_{\rm gas}$. Far left: [HCO$^+$]/[N$_2$H$^+$]. Middle left: [CO/C$^+$]. Middle right: [C/C$^+$]. Far right: Density-weighted average cosmic ray ionization rate, $<\zeta_\rho>$. Models in descending order from top: LDI, LRI, HDI}
\end{figure*}

Figure \ref{fig:ratsS} shows the same column density ratios for three models including the embedded protoclusters: LRI, LDI and HDI. Here the trends are significantly more complicated. The ${\rm [HCO^+/N_2H^+]}$ ratio still only varies by a factor of two throughout the parameter space, but it exhibits more complex behavior. The ratio decreases with $\Sigma_{\rm cl}$ and rises with $N_*$ up to some maximum, with an additional increase in ${\rm [HCO^+/N_2H^+]}$ for N$_* \approx 10^4$ for the LDI and HDI model. To understand this, we can look at the abundance profiles of the LDI model in detail in Figure \ref{fig:profVir}. The abundance of HCO$^+$ increases with both with $\Sigma_{\rm cl}$ and N$_*$ with the abundance profile flattening as a function of $A_V$ for $\Sigma_{\rm cl} > 1$ g cm$^{-2}$. For N$_2$H$^+$ the trends are separated by an A$_V$ threshold at $A_V = 1$. At $A_V < 1$, the abundance of N$_2$H$^+$ increases like HCO$^+$, with $\Sigma_{\rm cl}$ and N$_*$. For $1 < A_V < 100$, the abundance of N$_2$H$^+$ is sensitive primarily to N$_*$. In high ionization environments, CO will be destroyed in the creation of HCO$^+$ due to interactions with H$_3^+$. These environments will also produce N$_2$H$^+$ which destroys CO to create HCO$^+$. This is likely the main driving cause in the abundance profiles: there is a reduction of CO and N$_2$H$^+$ in the dense more ionized gas, and an systematic increase in HCO$^+$. The ${\rm [CO/C^+]}$ ratio increases monotonically across two orders of magnitude towards high $\Sigma_{\rm cl}$ and low $N_*$: cold gas is less ionized (lower right corner), so the amount of CO increases with respect to C$^+$.  ${\rm [C/C^+]}$ shows a different trend  compared to ${\rm [CO/C^+]}$. High ionization rates, in both the LDI and HDI models, have an increased ${\rm [C/C^+]}$ in lower mass clouds hosting smaller clusters and a decreased ${\rm [C/C^+]}$ at high $\Sigma_{\rm cl}$ compared to the LRI model. The ${\rm [C/C^+]}$ ratio is nearly flat across the $\Sigma_{\rm cl} - N_*$ parameter space in the HDI model. Clouds with fewer CRs and more gas shielding to the incident the FUV radiation have more C compared to C$^+$. 

\begin{figure*}
	\centering
    \begin{tabular}{c}
    \includegraphics[width=0.9\textwidth]{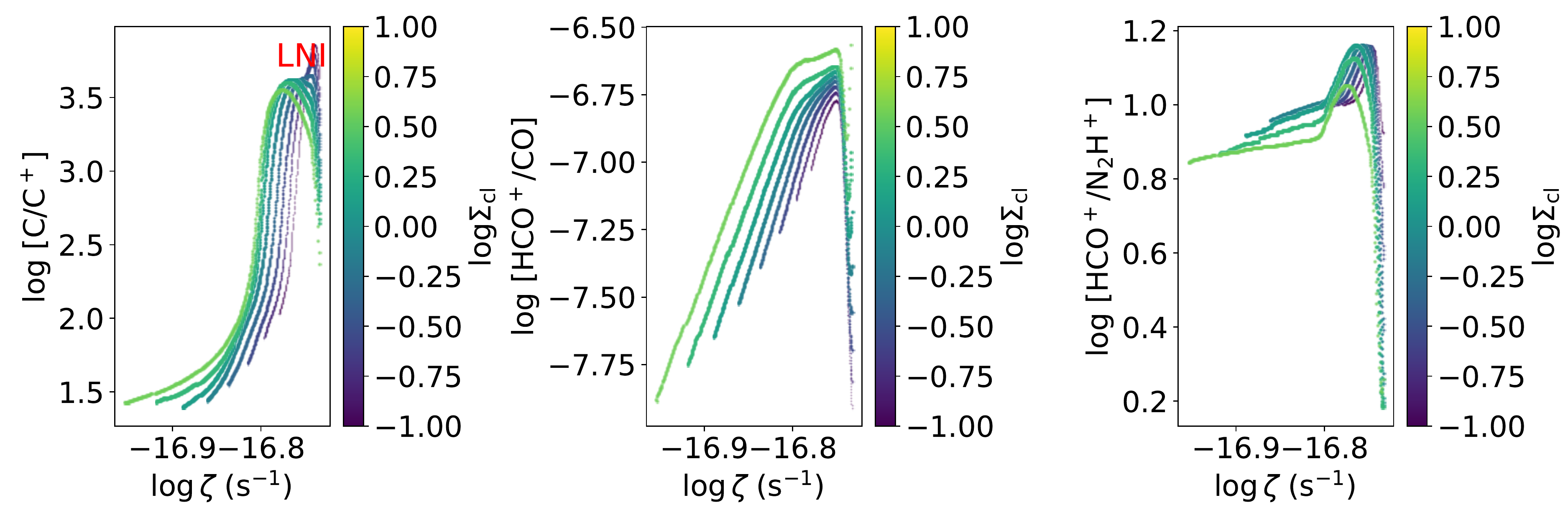} \\ \hline
    \includegraphics[width=0.9\textwidth]{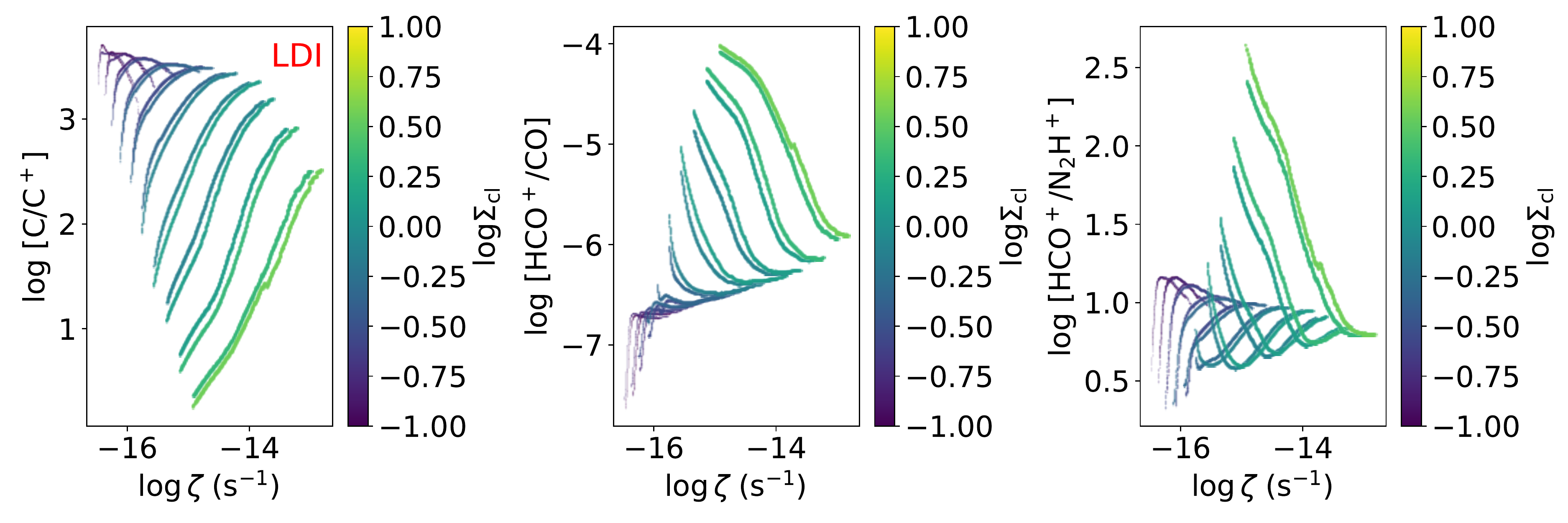} \\ \hline
    \includegraphics[width=0.9\textwidth]{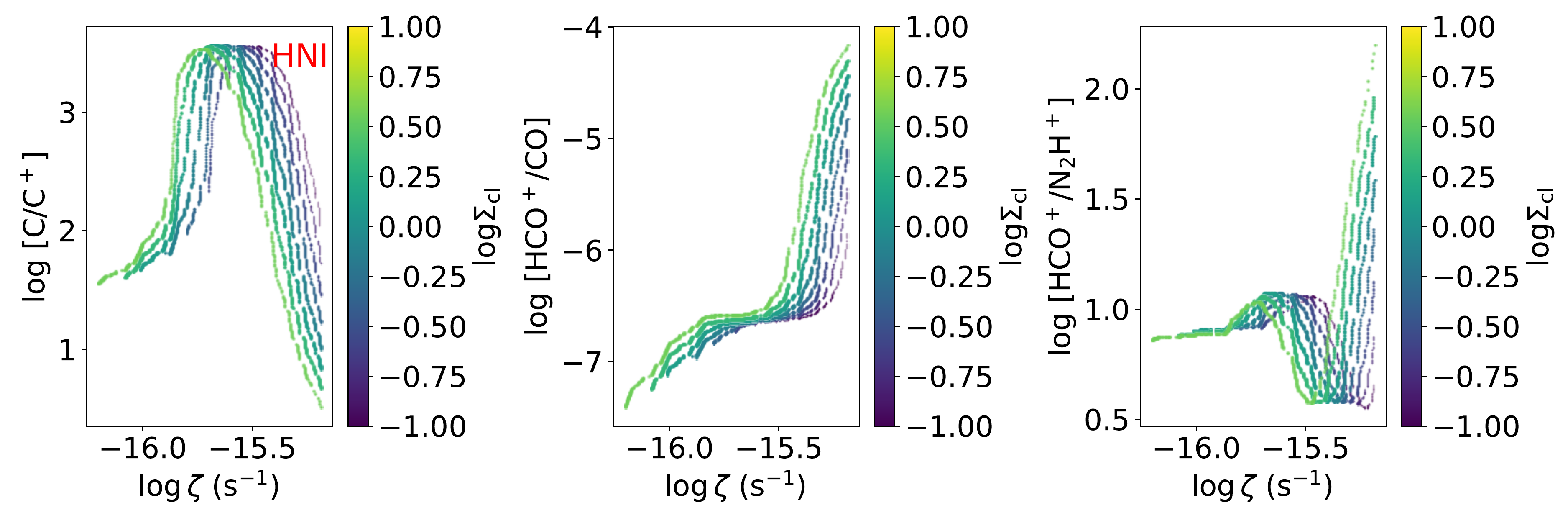} \\ \hline
    \includegraphics[width=0.9\textwidth]{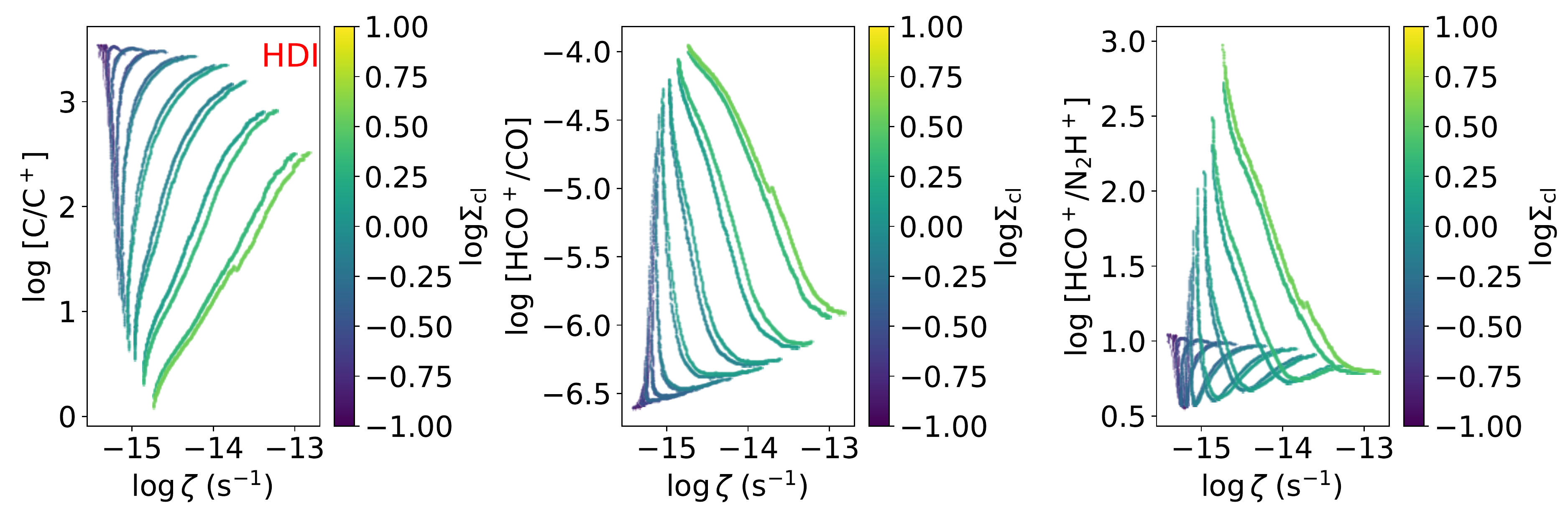} \\
    \end{tabular}
	\caption{\label{fig:zetaLocal} Cosmic ray ionization rate as a function of different abundance ratios. Left: C/C$^+$, middle: HCO$^+$/CO and right: HCO$^+$/N$_2$H$^+$. Models in descending order from top: LNI, LDI, HNI, HDI.}
\end{figure*}

Observational measurements of $\zeta$ in dense gas typically use astrochemical modeling and local abundance ratios (See \S\ref{sec:compcrir}). Figure \ref{fig:zetaLocal} plots the CRIR for models with $5\% \leq \varepsilon \leq 25\%$ as a function of different abundance ratios. A good CRIR indicator should exhibit a monotonic trend in response to changes in the CR flux. The LNI model does not exhibit much change in the CRIR, so the local trends depend on density and radiative effects. In the HNI model, only the ${\rm [HCO^+/CO]}$ ratio exhibits a monotonic trend.

The models with sources show completely different abundance ratios because the dense gas is warmer and the ionization rates are higher. In all of these cases, the ratios are monotonic for $\Sigma_{\rm cl} \gtrsim 1$ g cm$^{-2}$. For $\Sigma_{\rm cl} \lesssim 1$ g cm$^{-2}$ each exhibits a similar trend as in the NI model subsets. This demonstrates that these diagnostics only constrain regimes where the CRIR influences the chemistry more than radiative or other heating processes.

\section{Discussion}\label{sec:disc}
\subsection{Model Assumptions and Caveats}\label{sec:caveat}
Our models require a variety of assumptions. First and most importantly, the models are one-dimensional and we assume protostars are clustered in the center. In reality, protostars are distributed throughout molecular clouds. Furthermore, the density distribution of molecular clouds is set by turbulence and is not a purely radial distribution. However, our results will hold qualitatively for the molecular gas around young, dense embedded clusters in molecular clouds, such as the central cluster in $\rho$ Oph. Second, our chemical network does not include any gas-grain chemistry or freeze-out \citep[see][]{mcelroy2013}. Therefore, we over-predict the CO abundance in regions where $n > 10^4$ cm$^{-3}$ and $T \lessapprox 30$ K \citep{vandishoeck2014}. Comparing this criteria to the temperature profiles in Figure \ref{fig:profVir} shows the models with $\Sigma_{\rm cl} < 0.5$ g cm$^{-2}$ and where $1 \leq A_V \leq 10$ are below the freeze-out temperature. When embedded sources are included, the densest gas heats to temperatures $> 50$ K. These temperatures lead to desorption from the grains, producing gas-phase CO: any CO-ice that formed before star formation occured would be evaporated back into the gas phase \citep{jorgensen2013, jorgensen2015}.

We do not fully solve the cosmic ray transport equations or the acceleration dynamics of protons in protostellar accretion shocks. We use analytic approximations to describe the acceleration of CRs at the protostellar shock and the transport out of both the parent core and cloud. \cite{gaches2018b} explore the changes in the CRIR for different transport regimes, shock efficiencies and magnetic fields. Differences in the protostellar magnetic field changes the maximum energy of the accelerated CRs but has little effect on the CRIR. However, the CRIR scales nearly linearly with the shock efficiency. Our results assume  CR transport in the rectilinear regime through the core. More diffusive transport would produce higher temperatures at the surface of the core than observed.  The details of the transport depend both on the magnetic field morphology and on the coupling between the particles and the field.  In molecular clouds, turbulence is much stronger than in the cores allowing CRs to diffuse across magnetic field lines rather than streaming along them \citep{schlickeiser2002}. Conversely, if the particles are well-coupled their trajectories would follow the field lines, potentially producing asymmetries in the CR flux. The directionality imposed by the protostellar outflow could cause CR beaming in the outflow direction or simply advect the particles along with the outflow gas \citep{rodgers-lee2017}. We assume CRs transport from their parent cores through the clouds by parameterizing the radial scaling by either diffusive ($1/r$) or free-streaming ($1/r^2$). We do not fully solve the transport equations, which has yet to be done for CRs propagating out of molecular clouds from internal sources.

\subsection{Comparison to Observed CRIRs}\label{sec:compcrir}
\begin{figure}
    \centering
    \includegraphics[width=0.5\textwidth]{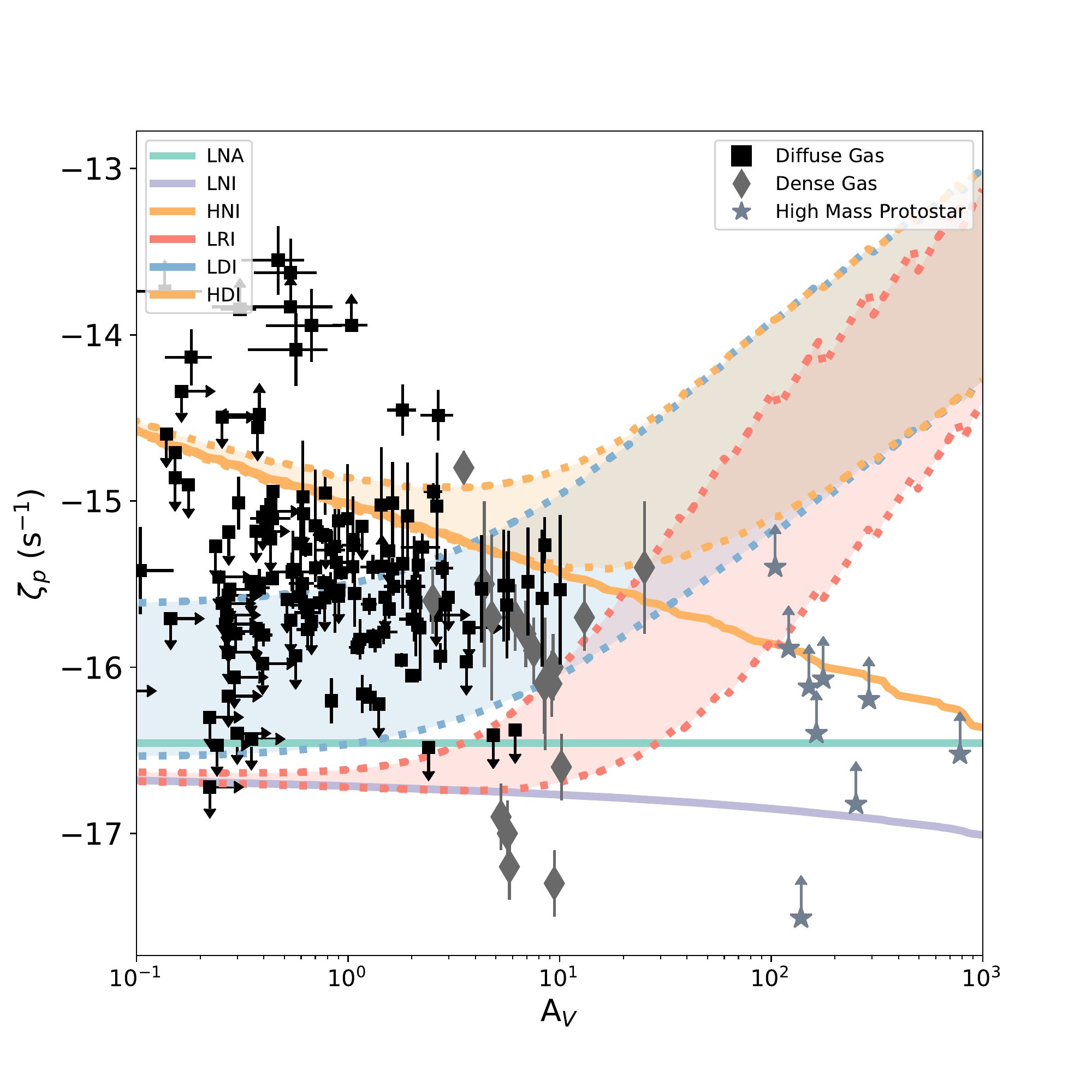}
    \caption{\label{fig:zetaav} Cosmic ray ionization rate, $\zeta$, versus $A_V$ for the six different models in Table \ref{tab:models}. The filled curves represent the $1\sigma$ spread from the models covering the ($\Sigma_{\rm cl}$ - $N_*$) parameter space. Squares represent diffuse gas measurements from \cite{indriolo2012} and \cite{indriolo2015}. Diamonds represent dense gas measurements from \cite{caselli1998}. Crosses represent observations towards high-mass protostars from \cite{boisanger1996, vandertak2000}.}
\end{figure}

Figure \ref{fig:zetaav} shows the results from the different PDR models in Table \ref{tab:models} compared to four different observational surveys covering a range of $1 \leq A_V \leq 10^3$. The CRIR is one of the trickier astrochemical parameters to constrain from observations. Unfortunately, no universal method is applicable to all clouds conditions. Historically, there have been two main methods: absorption measurements of simple ions, such as H$_3^+$ or OH$^+$, in the infrared or molecular line observations using key molecules in neutral-ion pathways along with astrochemical modeling. H$_3^+$ is typically  thought to be among the best tracers of the CRIR due to its simple chemistry. However, H$_3^+$ is only observed in infrared absorption, limiting its use to sight lines with bright background sources.  \cite{indriolo2012} and \cite{indriolo2015} used H$_3^+$, H$_2$O$^+$ and OH$^+$ absorption to trace the CRIR in diffuse gas with $A_V < 1$ and found the CRIR in low $A_V$ gas varies between $10^{-16}$ -$10^{-14}$ s$^{-1}$. The gas at low A$_V$ is particularly sensitive to external influences, motivating the  need to model the chemistry with external CR spectra derived from examining the local galactic environment. The grouping of points at low $A_V$ with high CRIR ($\zeta \geq 10^{-14}$ s$^{-1}$) are clouds in sightlines towards the galactic center and thus in environments with extreme external particle irradiation. \cite{caselli1998} used a combination of HCO$^+$, DCO$^+$ and CO together with analytic chemistry approximations to infer the CRIR in 24 dense cores. Their observations exhibit a nearly bi-model distribution: some  are clustered at $\zeta \lessapprox 10^{-17}\,{\rm s}^{-1}$, while the majority are at $\zeta \approx 10^{-16}\,{\rm s}^{-1}$. They infer the ionization rates using the abundance ratios of [DCO$^+$/HCO$^+$] and [HCO$^+$]/[CO] under 0D spatial assumptions and a reduced analytic chemical network. Finally, we include the CRIRs from the \cite{vandertak2000} survey towards single high-mass protostars with the central protostar being massive enough to provide a bright background source for H$_3^+$ absorption. They find CRIRs scattered from $10^{-17}$ to 10$^{-16}$ using an assumed H$_3^+$  abundance and density distribution. They find the observed H$_3^+$ column density increases with cloud distance,  which can be explained by contamination from low-density clouds along the line of sight. 

Our model results show good agreement with the inferred CRIRs from \cite{indriolo2012}, \cite{indriolo2015} and \cite{caselli1998}. We find the LDI model is able to replicate the spread in the CRIR. There are two main controlling factors for the CRIR in the clouds: the number of embedded sources and the cloud environment. Embedded sources create a natural dispersion in $\zeta$ for different molecular cloud masses and star formation efficiencies. Without internal sources, there is no spread in our modelled CRIR as a function of column density. In order to represent the observations, the external CRIR must be increased instead for different regions. Local sources of CRs, such as nearby OB associations or supernova, contribute significantly to the CR flux at the cloud boundary. As the external CR flux is increased, the impact of attenuation also increases due to the rapid reduction in low energy CRs. Figure \ref{fig:zetaav} shows that the impact of attenuation is different between the HNI and LNI models. For the LNI model, $\zeta$ changes by less than an order of magnitude across 4 orders of magnitude in $A_V$. Conversely, the HNI model CRIR decreases by 2 orders of magnitude due to an overall reduction in MeV-scale CRs. The HNI model over predicts the CRIRs measured in diffuse gas to CRIRs measured near high-mass protostars, excluding the galactic center sightlines. However, the $\mathcal{H}$ spectrum is the maximal CR spectrum from Voyager-1 observations \citep{stone2013,ivlev2015}. The LNI and LNA models under-predict the observed CRIR for all but a few sight-lines. Thus, Figure \ref{fig:zetaav} demonstrates that it is essential to consider the cloud environment and properly treat the CR physics and cloud density distributions. Models without attenuation only represent the CRIR within narrow ranges of $A_V$ and not in the cloud interiors. Figure \ref{fig:zetaav} also underscores that the low energy CR spectrum, which if often adopted in astrochemical modelling, is a poor fit to the majority of the observations.

\subsection{Challenges for Deriving the CRIR from Chemical Diagnostics}\label{sec:constrain}
There have been numerous attempts to find  chemical diagnostics that are strong tracers of the CRIR \citep{caselli1998, neufeld2010, neufeld2017, indriolo2007, indriolo2012, indriolo2015, albertsson2018}. Some of these, such as [DCO$^+$]/[HCO$^+$], cannot be modelled with the current {\sc 3d-pdr} version due to the lack of deuterium and isotopic chemistry. Most probes of the CRIR are based on the local abundance, which is difficult to directly ascertain from observations. The use of column density ratios typically assumes the line emission observed between species is co-spatial. Figures \ref{fig:ratsI} and \ref{fig:ratsS} examine effects of CR physics on the ${[\rm HCO^+/N_2H^+]}$, ${[\rm CO/C^+]}$ and ${\rm [C/C^+]}$ column density ratio diagnostics. However, we find that none of these ratios are monotonically
sensitive to the average CRIR, shown for the LDI model in Figure \ref{fig:colVir}. The [CO/C$^+$] column density ratio is anti-correlated with the density-averaged CRIR because the amount of CO declines while the amount of C$^+$ increases in large, massive clusters.

Local abundance ratios are used for fitting observations with astrochemical models which assume the ratio of the column densities is equal to the ratio of the abundances \citep{indriolo2012}. These are constrained through the use of 0-D spatial models, where a single density, temperature and extinction are evolved over time. However, this ignores the physical structure of clouds, which have non-uniform density, temperature, FUV and, as we show here, CRIR distributions \citep{clark2012,offner2013,gaches2015,glover2015,seifried2017}. As Figure \ref{fig:zetaLocal} shows, in models without internal sources, none of the abundance ratios are strong diagnostics. Furthermore, the range in the CRIR is small despite some large changes in each of the ratios. For models with internal sources, the CRs from embedded sources dominate the chemistry throughout the cloud. Here, we find the ratios of C/C$^+$ and HCO$^+$/CO are mostly monotonic with the CRIR. However, there is significant variation with $\Sigma_{\rm cl}$ and thus with the gas mass. The results signify that more careful physical and chemical modeling needs to be done to accurately constrain the CRIR in high-surface density, star-forming regions. The ratios are only a good diagnostic in regions where CRs dominate the thermo-chemistry.

Recently, \cite{lepetit2016} used H$_3^+$ absorption to infer the CRIR and physical conditions in the CMZ. They used a similar relation to Equation \ref{eq:ebal}:
\begin{equation}
    N({\rm H_3^+}) = 0.96\frac{\zeta L}{k_e}\frac{f}{2x_e},
\end{equation}
where $k_e$, $f$, and $x_e$ have the same definition as in Equations \ref{eq:ebal} and \ref{eq:simpanaly} and $L$ is the size of the cloud. They fit observed H$_3^+$ column densities with PDR models as a function of the gas density, $n_H$, and the size of the cloud, $L$. Most sightlines are well fit using their method by clouds with densities $10 \leq n_H \leq 100$ cm$^{-3}$ corresponding to $5 \lesssim L \lesssim 100$ pc. Our models suggest that these length scales would incur CR screening effects which would change the CRIR. Similarly, for the high density clouds, the energy losses will deplete MeV CRs and reduce the CRIR. In these cases, there is a further degeneracy in the $n_H - L$ plane resulting in an average decrease in the CRIR. The reduction would systematically produce model fits with lower densities to correct for the lower CRIR. 

\cite{rimmer2012} used a similar hybrid approach, adopting a prescription for  $\zeta(N)$ ad-hoc with the {\sc Meudon} PDR code \cite{lepetit2006} to model the Horsehead Nebula. They found their high $\zeta(N_H)$ model improved agreement over standard constant CRIR  PDR prescriptions. However, their treatment of $\zeta(N)$ is static and fixed in time. The decoupling of CR attenuation and chemistry is only a good approximation if the abundance of neutral Hydrogen (H, H$_2$) does not change much in time, ensuring that the CR spectrum is constant in time. The new approach presented here will allow $\zeta(N)$ to be connected to the chemical time evolution.

\subsection{Impact of Cosmic Ray Feedback on Cloud Chemistry}\label{sec:chemImpact}
The {\it Herschel} Galactic Observations of Terahertz C+ (GOT C+) \citep{pineda2013} survey mapped [C II] 158 $\mu$m emission over the whole galactic disk, providing the best constraint on where [C II] emission originates. \cite{pineda2013} found that nearly half of the [C II] emission originates from dense photon dominated regions with about another quarter of the emission from CO-dark H$_2$ gas. \cite{clark2019} performed synthetic observations of young simulated molecular clouds and found the majority of their [C II] emission originates from atomic-Hydrogen dominated gas. This discrepancy was explained by the time evolution of molecular gas due to star formation and feedback. The results presented here provide an complementary explanation for the [C II]-bright molecular gas. When protoclusters become active, they accelerate CRs into the densest regions of molecular clouds. Figure \ref{fig:profVir} shows that high-mass protoclusters will lead to [C II]-bright H$_2$ dominated gas since CRs i) increase the gas temperature to values closer to the [C II] excitation temperature of 91.2 K, ii) increase the abundance of C$^+$ in dense gas due to the destruction of CO and iii) do not significantly alter the abundance of H$_2$. 

In local star-forming regions, the lowest inversion transitions of ammonia, NH$_3$, have been widely used to map the dense gas cores within molecular clouds \citep[i.e.,][]{goodman1993, jijina1999, rosolowsky2008, wienen2012}. {\rm Ammonia remains optically thin, and while it does suffer from depletion, it's formation is enhanced in regions where CO freezes out \citep{caselli2012} (although this effect is not included in our models)}. However, this also makes ammonia much more susceptible to local variations in the FUV radiation field, temperature and CRIR. Recently, the Green Bank Ammonia Survey (GAS) mapped all the Gould Belt clouds with $A_V > 7 \, {\rm mag}$ \citep{friesen2017}. The DR1 data show the line-of-sight averaged abundance, $X({\rm NH_3}) = N({\rm NH_3})/N({\rm H_2})$, exhibits a spread through molecular clouds. The spread could be caused by the porosity of molecular clouds allowing more FUV radiation into regions of dense gas. However, our 1D models also exhibit a variation in this abundance measurement for the models with internal sources (LRI, LDI and HDI). Figure \ref{fig:profVir} shows that ammonia is depleted in clusters exhibiting more embedded star formation by a couple of orders of magnitude. The abundance within the dense gas goes from 10$^{-8}$ in small clusters to 10$^{-10}$ in the largest. Furthermore, the gas also heats up leading to stronger emission in higher transitions, such as NH$_3$(3,3). \cite{redaelli2017} examined the NH$_3$ GAS map of the Barnard 59 clump in more detail. They found that the abundance appears to drop in gas around the central central 0/1 protostar. The dust temperature shows a clear increase around the same source with a slight increase in the ammonia excitation temperature. 

\subsection{Impact of Cosmic Ray Feedback on Chemistry in Dense Cores}
Protostars are observed to be dimmer than classic collapse models predict, i.e.,  the ``Protostellar Luminosity Problem" \citep{dunham2014}. One possible solution is that accretion is strongly episodic \citep{kenyon90,vorobyov09,offner2011}. Although  our models assume steady-state accretion, we can infer the impact of large bursts of accretion on cloud chemistry. An accretion burst leads to a stronger accretion shock, which in turn produces higher energy CRs and a higher CRIR. The CRIR increase in the dense gas then raises the temperature. The higher temperatures, whether caused by radiative or CR heating, lead to several different chemical effects. First, molecules frozen onto dust grains will evaporate, both by thermal desorption \citep{oberg2016} and CR-induced desorption \cite{hasegawa1993}, into the gas phase. Second, the increase of the CRIR will increase the ionization fraction leading to a chain of ion-neutral reactions following H$_3^+$. Finally, the elevated radiative and CR flux may be strong enough to destroy some molecules. \cite{jorgensen2015} and \cite{frimann2016} showed that episodic accretion can cause the sublimation of CO-ice and explain the excess C$^{18}$O emission observed near protostars. Intuitively, a burst a CRs will lead to a reaction chain: H$_3^+$ is created, thereby leading to the destruction of CO to form HCO$^+$. However, the increase in CRs will provide a large population of free electrons which recombine with HCO$^+$ to form CO. HCO$^+$ also interacts with water and other dipole neutrals (in the case of water, the reaction leads to the formation of CO and protonated water). HCO$^+$ is observed to be depleted near protostars that have undergone episodic accretion \citep{jorgensen2013}. Ices sublimated by an accretion burst will cause  a more active gas-phase chemistry and lead to an increase in carbon-chain molecules in molecular clouds as well as increase gas-phase CO in the dense gas where it would otherwise freeze-out. Overall, the addition of CRs magnifies the effect of an accretion burst. Temperatures increase beyond that expected from radiative heating alone. This suggests that a smaller change in accretion rate may be needed to produce the observed chemical changes.

\subsection{Implications for Comparing Data and Models}
Synthetic observations of hydro-dynamic simulations are a vital tool for comparing theoretical predictions to observations. The synthetic observations may treat the chemistry in different ways: from assuming a constant abundance of some molecule to post-processing simulations with an astrochemical code or using the reduced-network chemistry from the hydrodynamic simulation \citep[see review by][]{haworth2018}. These synthetic observations are used to gauge how well the simulations correspond to the observed universe. As such, it is paramount to ensure that all astrochemical parameters are as  accurate as possible. Radiation-MHD simulations can provide the density and temperature at every point \citep[e.g.,][]{Offner2009}. Simulations also now often include FUV radiation and optical depth calculations using packages such as {\sc Fervent} \citep{baczynski2015}, {\sc TreeRay} (W\"{u}nsch et al., in preparation, used in \cite{haid2019}) and {\sc Harm$^2$} \citep{rosen2017}. These methods can provide the FUV radiation and/or optical depth into the cloud. Our results, here, show that the H$_2$ optical depth into the cloud should be considered when calculating the appropriate CRIR for post-processing. Typically, the CRIR is held constant throughout the entire simulation domain, which will lead to systematic differences in the simulation line emission.

\section{Conclusions}
We implement cosmic-ray attenuation in the public astrochemistry code {\sc 3d-pdr}. The implementation uses the H$_2$ column density from the chemistry to attenuate the CR spectra. We couple the code to the protostellar CR models from \cite{gaches2018b}, which solve for the total attenuated protocluster CR spectrum as a function of the cloud surface density, $\Sigma_{\rm cl}$, and number of constituent protostars, N$_*$. We present one-dimensional astrochemical models for molecular clouds with a wide range of $\Sigma_{\rm cl}$ and $N_*$.  We compare the abundance distributions for a low external CR spectrum, representing an extrapolation of the Voyager-1 data,  and a high external CR spectrum, representing a maximal correction for solar influences. Our model results show that CRs originating from the accretion shocks of protostars affect the chemistry of the surrounding molecular cloud. We conclude the following:
\begin{itemize}
    \item Models with no sources or attenuation cannot explain observed CRIRs. Models with no internal sources but a higher ($\mathcal{H}$) external spectrum (HNI) match the observed CRIRs, although it may under-predict the CRIRs inferred for high-mass protostars. We find that a model using the commonly adopted spectrum with internal sources (LDI) matches both the low and mid $A_V$ observations of $\zeta$ and the observed spread.
    \item CRs accelerated by protostellar accretion shocks significantly alter the Carbon chemistry in star-forming clouds. The amount of neutral and ionized Carbon increases in the dense gas as the number of protostars increases. Models with embedded sources (LDI, LRI, HDI) increase the amount of C, HCO$^+$ and NH$_3$ at lower $A_V$ and decrease the abundance of CO and NH$_3$ at higher $A_V$. Overall, models including internal sources (LDI, LRI and HDI) exhibit a higher abundance of HCO$^+$ and H$_3^+$ with $\Sigma_{\rm cl}$ and $N_*$.
    \item Approximations that use H$_3^+$ and C-based tracers to estimate the CRIR systematically under-predict the CRIR unless CRs are the dominant source of ionization. The Reduced Analytic Approximation, which uses the abundances of H$_3^+$, CO and O,always produces more accurate values of the CRIR,  highlighting the importance of obtaining accurate Oxygen and Carbon Monoxide abundances within molecular clouds. 
    \item Ions are systematically under produced using the canonical CRIR while CO is over produced. Internal sources created a dispersion in the distribution of column densities by driving more active ion-neutral chemistry deep within molecular clouds.
    \item Models using the low external CR spectrum  ($\mathcal{L}$) and/or no internal sources of CRs under estimate the H$_3^+$ column density by a order of magnitude or more. 
    \item Internally-accelerated CRs will naturally lead to molecular gas which become CO-deficient but [C II]-bright, particularly for high surface density molecular clouds hosting large clusters.
    \item Including CR attenuation in PDR models will help break the denegeracy in astrochemical modeling between the density, CRIR and FUV radiation.
\end{itemize}

As protoclusters grow in constituent nubmers, the impact on the chemistry is amplified, greatly so if CRs diffuse out of molecular clouds rather than stream. Comparison to observed CRIRs suggest the external CR spectrum, attenuation and internal sources are important for modelling the chemistry of molecular clouds. However, the current uncertainties are large due to lack of observational data that can simultaneously constrain the density, FUV radiation and CRIR on molecular cloud scales. Observations to constrain the CRIR within dense gas necessitate multi-line data, to independently  determine the temperature as in e.g., \cite{ceccarelli2014}, and multi-species data, to act as astrochemical diagnostics as in e.g., \cite{caselli1998}. The {\sc 3d-pdr} CR attenuation {\sc Fortran} module can be included in any {\sc Fortran} astrochemistry code. 

\facility{Massachusetts Green High Performance Computing Center}
\software{ \begin{itemize} \item {\sc 3d-pdr} \citep{bisbas2012} 
\item {\sc matplotlib} \citep{matplotlib2007} 
\item {\sc NumPy} \citep{numpy2011} 
\item {\sc SciPy} \citep{scipy2001} 
\item {\sc JupyterLab} \end{itemize}}

\acknowledgments
SSRO and BALG acknowledge support from the National Science Foundation (NSF) grant AST-1510021. SSRO was also supported by NSF CAREER grant AST-1650486. TGB acknowledges funding by the German Science Foundation (DFG) via the Collaborative Research Center SFB 956 ``Conditions and impact of star formation''. The authors thank helpful discussions with Neal Evans and Nick Indriolo and the anonymous referee for their useful comments. The calculations performed for this work were done on the Massachusetts Green High Performance Computing Center (MGHPCC) in Holyoke, Massachusetts supported by the University of Massachusetts, Boston University, Harvard University, MIT, Northeastern University and the Commonwealth of Massachusetts.

\bibliography{crchem1}
\end{document}